\documentclass[a4paper,11pt]{article}
\usepackage{rotating}
\usepackage{jheppub} 
\usepackage[utf8]{inputenc}
\usepackage[T1]{fontenc} 
\usepackage{placeins} 
\usepackage{chngpage}
\usepackage{amsmath,amsfonts}
\usepackage{graphicx}
\usepackage{color}
\usepackage{xcolor}
\usepackage{relsize}
\usepackage{epstopdf}
\usepackage{hyperref}
\usepackage{mathrsfs}
\usepackage{ragged2e}
\usepackage{amssymb}
\usepackage{placeins}
\usepackage[normalem]{ ulem }
\usepackage{amsthm}
\usepackage{comment}
\usepackage{graphics}	
\usepackage{caption}
\usepackage{subcaption}
\usepackage{verbatim}
\usepackage{arydshln}
\usepackage{slashed}
\usepackage{tikz}
\usepackage{xspace}

\usepackage{tabularx,booktabs}
\usepackage{multicol}
\usepackage{array,multirow}
\usepackage{booktabs}
\usepackage{colortbl}
\usepackage{float}
\usepackage{xspace}

\usepackage{tikz}
\usepackage[compat=1.1.0]{tikz-feynman}

\newcommand{\rivet}{R\protect\scalebox{0.8}{IVET}\xspace}

\newcommand{\contur}{\textsc{Contur}\xspace}

\newcommand{\madgraph}{\textsc{Madgraph5}\xspace}

\newcommand{\cgeff}{\ensuremath{c^{\text{eff}}_{agg}}\xspace}
\newcommand{\cteff}{\ensuremath{c^{\text{eff}}_{t}}\xspace}
\newcommand{\cfeff}{\ensuremath{c^{\text{eff}}_{f}}\xspace}
\newcommand{\cbeff}{\ensuremath{c^{\text{eff}}_{b}}\xspace}
\newcommand{\cto}{\ensuremath{c_t^0}\xspace}
\newcommand{\cgo}{\ensuremath{c_{\tilde{G}}^0}\xspace}
\newcommand{\ttbar}{\ensuremath{t\bar{t}}\xspace}
\newcommand{\pt}{\ensuremath{p_{\textrm{T}}}\xspace}
\newcommand{\MET}{\ensuremath{E_T^{\rm miss}}}

\renewcommand{\to}{\rightarrow}

\newcommand{\ETmiss}{E_{\text{T}}^{\text{miss}}}

\newcommand{\mf}{\text{f}}
\newcommand{\mfp}{\text{f'}}
\newcommand{\be}{\begin{equation}}
\newcommand{\ee}{\end{equation}}
\makeatletter
\gdef\@fpheader{}
\vspace*{0cm}
\makeatother

\begin{document}

\title{Probing the coupling of axions to tops and gluons with LHC measurements}

\author[a]{Jon Butterworth,}
\author[a]{Matthew Cullingworth,}
\author[a]{Joseph Egan,} 
\author[b]{Fabian Esser,}
\author[c]{Veronica Sanz}
\author[d]{ and Maria Ubiali}
\affiliation[a]{Department of Physics \& Astronomy, University College London, London, WC1E 6BT, United Kingdom}
\affiliation[b]{Institute of Particle and Nuclear Physics (IPNP), Faculty of Mathematics and Physics, Charles University
Prague, V Holešovičkách 2, 180 00 Praha 8, Czech Republic}
\affiliation[c]{Instituto de F\'isica Corpuscular (IFIC), Universidad de Valencia-CSIC, E-46980 Valencia, Spain}
\affiliation[d]{DAMTP, University of Cambridge, Wilberforce Road, Cambridge CB3 0WA, UK }

\emailAdd{J.Butterworth@ucl.ac.uk}
\emailAdd{joe.egan.23@ucl.ac.uk}
\emailAdd{esser@ific.uv.es}
\emailAdd{veronica.sanz@uv.es}
\emailAdd{M.Ubiali@damtp.cam.ac.uk}

\date{\today}

\abstract{We study axion-like particles (ALPs) whose dominant interactions are with gluons and third-generation quarks, and whose couplings to light Standard Model (SM) particles arise at one loop. These loop-induced effects lead to ALP decays and production channels that can be probed at the LHC, even when tree-level couplings are absent. Using an effective field theory (EFT) description that includes momentum-dependent corrections from radiative effects, we reinterpret a wide range of LHC measurements via the \contur framework to derive model-independent constraints on the ALP parameter space.
We show that LHC data place meaningful bounds in the plane of effective couplings $\cto/f_a$ and $\cgo/f_a$, and that these limits are sensitive to the  UV origin of the ALP-top and ALP-gluon couplings. We discuss representative scenarios where either \cto or \cgo vanishes at the matching scale, and highlight the role of EFT running and mixing in generating observable signals. We also assess the domain of validity of the EFT approach by comparing the typical momentum transfer $\sqrt{\hat s}$ in sensitive regions to the underlying scale $f_a$.
Our results demonstrate the power of loop-aware EFT reinterpretation of SM measurements in probing otherwise elusive ALP scenarios. The framework presented here can be readily extended to include couplings to other fermions and to accommodate ALP decay or long-lived signatures.
}

\keywords{Axion-like particles, top quark,  LHC physics}

\maketitle

\section{Introduction}
\label{sec:intro}

Originally motivated by the QCD CP problem~\cite{Peccei:1977hh,Peccei:1977ur,Wilczek:1977pj,Weinberg:1977ma}, axion-like particles (ALPs)
appear in many extensions of the Standard Model (SM), see Ref.~\cite{Choi:2020rgn} for a recent review and~\cite{AxionLimits} for a 
public repository showing many of the available constraints on the ALPs parameters from a variety of direct and indirect searches 
in the field of high-energy physics, nuclear and atomic physics, cosmology and astrophysics.

ALPs can couple to any SM particle and thus lead to novel signatures at colliders over a broad range of masses. 
Phenomenological studies of the ALP coupling to gluons~\cite{Mimasu:2014nea}, diboson pairs~\cite{Jaeckel:2015jla,Brivio:2017ije,Bauer:2017ris,Craig:2018kne},
top pair~\cite{Esser:2023fdo,Bisal:2025jwv} and di-Higgs~\cite{Esser:2024pnc} using collider probes have shown that sensitivity to ALPs exists at
colliders, not only through resonant signatures but also via non-resonant production of a light
ALP~\cite{Gavela:2019cmq,No:2015bsn,Carra:2021ycg}.  
Such studies motivated several experimental analyses at the LHC~\cite{Alimena:2019zri,CMS:2018erd,ATLAS:2022abz,ATLAS:2023ofo,ATLAS:2023ian,ATLAS:2023zfc,ATLAS:2023etl,ATLAS:2024vpj,CMS:2024ulc}.

On the other hand, there are important gaps, in particular in collider studies of the ALP coupling to gluons.  
While studies exist for high-mass ALPs~\cite{Arganda:2018cuz}, and expected sensitivities
have been proposed for example in Ref.~\cite{Haghighat:2020nuh}, both the ALP-gluon coupling for low-mass ALPs,
and its interplay with the ALP-top coupling, have not been studied in detail with real experimental data.
This paper aims to cover this gap and to motivate further collider studies in this direction. 

We focus on an ALP that couples predominantly to the top quark and to gluons. Such a coupling structure is well motivated both theoretically 
and phenomenologically. The top quark, being the heaviest SM fermion, is expected to play a special role in the UV completion of the SM, and its interactions 
with new pseudoscalars can be naturally enhanced in models with partial compositeness~\cite{Redi:2011zi, Matsedonskyi:2012ym, Pomarol:2012qf}. Simultaneously, 
ALP couplings to gluons are often generated through anomalies or heavy coloured fermion loops.

We adopt a simplified effective field theory (EFT) approach in which the ALP couples only to the top quark and gluons at tree level. This minimal setup 
allows us to isolate and study the interplay of these two couplings, which in turn radiatively induce all other effective interactions with SM particles. 
These loop-induced effects are particularly important for the phenomenology at colliders, where processes such as ALP-mediated top pair or dijet production provide powerful probes.

To do this, we make use of the \texttt{FeynRules}~\cite{Alloul:2013bka} model of Ref.~\cite{Brivio:2017ije}, translated 
as a UFO~\cite{Degrande:2011ua} and read by \madgraph\cite{Alwall:2014hca}, to simulate ALP production at the LHC in 
proton-proton collisions at 13~TeV.
The simulated events are then passed through \rivet~4~\cite{Buckley:2010ar,Bierlich:2024vqo} and analysed with
\contur~3.1~\cite{Butterworth:2016sqg,Buckley:2021neu,CONTUR:2025yis}.
This involves injecting the ALP contribution on top of the SM predictions for a wide range
of LHC measurements, and evaluating the ratio between the likelihood evaluated for the SM alone given the data,
versus the likelihood for SM+ALP contributions, to obtain expected and actual exclusion contours.
We derive exclusion limits in the plane of effective ALP couplings, both at tree level and after including 
loop effects, and study their dependence on the ALP mass and the assumptions about the validity of the EFT. 
We also explore concrete scenarios in which the ALP arises as a pseudo-Goldstone boson, and demonstrate how 
collider data can provide insight into the UV structure of the theory, including the presence or absence of anomalous gluonic couplings.

The paper is structured as follows. In Sec.~\ref{sec:theory}, we introduce the theoretical framework, 
including the effective couplings of the ALP to SM fields and their radiative structure. In 
Sec.~\ref{sec:models}, we discuss model-building motivations and provide an explicit realization within a 
composite Higgs framework. Sec.~\ref{sec-results} presents our results from the reinterpretation of LHC measurements. 
In Sec.~\ref{sec:validity}, we examine the domain of validity of the EFT used in the analysis. We conclude in Sec.~\ref{sec:concls} 
with a discussion of the implications and outlook.

\section{Theoretical framework}
\label{sec:theory}
Axion-like particles (ALPs) are pseudo-Nambu–Goldstone bosons associated with the spontaneous breaking of approximate global symmetries. 
As pseudoscalars, ALPs are CP-odd, and their couplings are constrained by a shift symmetry,
\( a \to a + C \), where \( C \) is a constant. This symmetry suppresses or forbids certain interactions, particularly those involving explicit mass terms or scalar potentials.

We consider scenarios in which the ALP is associated with a high symmetry-breaking scale \( f_a \gg v \), where \( v \) is the electroweak scale. 
In this regime, the ALP interactions can be systematically described using an effective field theory (EFT) expansion, organised in powers of \( 1/f_a \).

The goal of this study is to investigate the couplings of ALPs to the top quark and to gluons. Focusing on the top quark is motivated by the 
fact that ALP–fermion interactions, when rewritten using the equations of motion, are proportional to the fermion mass. As a result, couplings 
to heavier fermions are expected to dominate, with the top quark playing a central role. Moreover, a preferential coupling to gluons would be expected if the origin of the ALP is tied to an ultraviolet theory with couplings to the SM strong sector.

Focusing on couplings to third-generation quarks, the leading interactions take the form
\begin{equation}
\label{eq:top_coupling_su2}
\mathcal{L} = \frac{\partial_\mu a}{f_a} \left( 
c_{Q_3}  \bar{Q}^3_L \gamma^\mu Q^3_L + 
c_{t_R} \bar{t}_R \gamma^\mu t_R + 
c_{b_R} \bar{b}_R \gamma^\mu b_R 
\right) ,
\end{equation}
where \( Q^3_L = (t_L, b_L)^T \) is the third-generation left-handed quark doublet, and \( c_{Q_3} \), \( c_{t_R} \), and \( c_{b_R} \) are model-dependent Wilson coefficients.
The couplings in Eq.~\eqref{eq:top_coupling_su2} are flavour-diagonal and respect the SU(2)\(_L\) gauge structure of the SM. In this work, we assume that the ALP does not mediate new sources of flavour violation\footnote{For studies exploring ALPs with non-trivial flavour structure, see Refs.~\cite{Bonilla:2022qgm,Carmona:2022jid,Carmona:2021seb,Chala:2020wvs,Bauer:2019gfk,Bauer:2021mvw}.}. Specifically, we consider the following axial-vector interaction between the ALP and the top quark,
\begin{equation}
\label{eq:top_coupling}
\mathcal{L} = 
  \cto \,  \frac{\partial_\mu a}{2 f_a} \,  (\bar t \gamma^\mu \gamma^5 t) \ ,
\end{equation}
where \( \cto = c_{t_{R}} - c_{Q_{3}} \) and \( t = (t_L, t_R)^T \). The superscript ``0'' denotes the tree-level value of the coupling, distinguishing it from loop-induced contributions that will be discussed later.

Using the equations of motion, this operator can be rewritten as a pseudoscalar coupling:
\begin{equation}
\mathcal{L} = 
-i \, \cto \,  \frac{m_t a}{f_a} \,  (\bar t  \gamma^5 t) \ ,
\end{equation}
which explicitly exhibits the proportionality to the top mass \( m_t \). This mass dependence reflects a generic feature of ALP couplings to fermions, implying a hierarchical structure aligned with the fermion mass spectrum. The top quark, being the heaviest SM fermion, is thus expected to have the strongest coupling, providing strong motivation for focusing on this interaction.

ALP couplings to SM gauge bosons also appear at leading order in the EFT expansion. In the unbroken phase, these are described by the dimension-5 operators:
\begin{equation}
\label{eq:Lgauge_unbroken}
\mathcal{L} =  - \frac{a}{f_a} \left( \cgo \, G_{\mu\nu} \tilde{G}^{\mu\nu} + c_{\tilde{W}}^0 \, W^I_{\mu\nu} \tilde{W}^{I,\mu\nu} + c_{\tilde{B}}^0 \, B_{\mu\nu} \tilde{B}^{\mu\nu} \right) \ ,
\end{equation}
where an implicit sum over colour indices is understood in the gluon term.

After electroweak symmetry breaking (EWSB), these interactions translate into couplings to the mass eigenstates of the gauge bosons:
\begin{align}
\label{eq:Lgauge_broken}
\mathcal{L}  - \frac{a}{f_a} \Big(
& c_{agg} \, G_{\mu\nu} \tilde{G}^{\mu\nu}
+ c_{aWW} \, W_{\mu\nu} \tilde{W}^{\mu\nu}
+ c_{aZZ} \, Z_{\mu\nu} \tilde{Z}^{\mu\nu} \nonumber \\
&+ c_{a\gamma Z} \, F_{\mu\nu} \tilde{Z}^{\mu\nu}
+ c_{a\gamma\gamma} \, F_{\mu\nu} \tilde{F}^{\mu\nu}
\Big) \ ,
\end{align}
where, at tree level, the coefficients are related to the unbroken-phase parameters by\footnote{Note that the couplings \( g_{aXX} \) used in Ref.~\cite{Bonilla:2021ufe} are related to those in Eq.~\eqref{eq:caXX} by, 
$c_{aXX} = \frac{f_a}{4} \, g_{aXX}$}
\begin{align}
c_{agg} &= \cgo, \\
c_{aWW} &= c_{\tilde{W}}^0, \\
c_{a\gamma\gamma} &= s_w^2 \, c_{\tilde{W}}^0 + c_w^2 \, c_{\tilde{B}}^0, \\
c_{aZZ} &= c_w^2 \, c_{\tilde{W}}^0 + s_w^2 \, c_{\tilde{B}}^0, \\
c_{a\gamma Z} &= 2 s_w c_w \, (c_{\tilde{W}}^0 - c_{\tilde{B}}^0),
\label{eq:caXX}
\end{align}
with \( s_w = \sin\theta_W \) and \( c_w = \cos\theta_W \).

\begin{figure}[t!]
    \centering
    \includegraphics[scale=1]{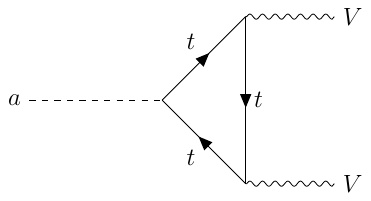}
    \includegraphics[scale=1]{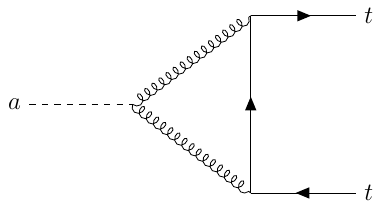}
    \caption{Loop-induced ALP interactions: (a) anomalous coupling to gauge bosons via a top loop, (b) induced ALP--top coupling from the gluon coupling}
    \label{fig:alp-loops}
\end{figure}

Throughout this work, we focus on the simplified scenario in which all ALP couplings to SM fields other than the top quark and the gluon are set to zero at tree level. This minimal setup allows us to isolate and quantify the impact of the \( a t \bar{t} \) and \( a gg \) couplings on LHC observables, while taking into account loop-induced effects on the remaining interactions.

These loop-induced couplings arise from diagrams involving the top quark and gluons, as illustrated in Fig.~\ref{fig:alp-loops}, where both the ALP–gauge boson vertex and the ALP–top vertex, or to any other fermion, can be generated radiatively.

\subsection{ALP couplings to vector bosons}
\label{sec:aVV}

Although only two ALP couplings are taken to be non-zero at tree level—\( \cto \) to the top quark and \( \cgo \) to gluons—radiative corrections generate additional effective interactions with electroweak gauge bosons. These loop-induced contributions have been computed explicitly in Ref.~\cite{Bonilla:2021ufe}.

In particular, the axial ALP–top coupling \( \cto \) induces one-loop corrections to the ALP– gauge boson couplings in Eq.~\eqref{eq:Lgauge_broken}, modifying the tree-level relations in Eq.~\eqref{eq:caXX}. When only \( \cto \) and \( \cgo \) are non-zero at tree level, the effective couplings become
\begin{align}
    \cgeff &= \cgo - \frac{\alpha_s}{8 \pi} \cto \, B_1 \left( \frac{4 m_t^2}{p^2} \right), \label{eq:cagg_1L}\\
    c_{a\gamma\gamma}^{\rm eff} &= -\frac{\alpha_{\rm em}}{3 \pi} \cto \, B_1 \left( \frac{4 m_t^2}{p^2} \right), \label{eq:cagammagamma_1L}\\
    c_{a\gamma Z}^{\rm eff} &= \frac{\alpha_{\rm em}}{4 \pi c_w s_w} \cto \, A^t(p^2), \label{eq:cagammaZ_1L} \\
    c_{aZZ}^{\rm eff} &= \frac{\alpha_{\rm em}}{4 \pi c_w^2 s_w^2} \cto \, B^t(p^2), \label{eq:caZZ_1L} \\
    c_{aWW}^{\rm eff} &= \frac{\alpha_{\rm em}}{4 \pi s_w^2} \cto \, C^t(p^2). \label{eq:caWW_1L}
\end{align}
Here $p$ denotes the 4-momentum of the ALP and $\sqrt{p^2}$ can be interpreted as the energy it carries~\cite{Bonilla:2021ufe}.
The loop functions $B_1$, $A^{\text{t}}$, $B^{\text{t}}$ and $C^{\text{t}}$ are given in App.~\ref{sec:appendix} in Eq.\ (\ref{eq:B1}), (\ref{eq:Af}), (\ref{eq:Bf}) and (\ref{eq:Cf}), respectively. 
In the high-energy limit \( p^2 \gg m_t^2, M_Z^2, M_W^2 \), the loop functions approach constant values: \( B_1(4 m_t^2/p^2) \to 1 \), \( A^t(p^2) \to \frac{8}{3} s_w^2 \), \( B^t(p^2) \to -\frac{4}{3} s_w^4 \), and \( C^t(p^2) \to 0 \), yielding simplified expressions\footnote{In the UFO model implementation\cite{Brivio:2017ije}, 
these effective electroweak couplings can be mimicked at high energy by setting \( c_{\tilde{B}} = - \frac{\alpha_{\rm em}}{3 \pi c_w^2} \cto \) and \( c_{\tilde{W}} = 0 \).},
\begin{align}
    \cgeff &= \cgo - \frac{\alpha_s}{8 \pi} \cto \label{eq:cgctsimple} \\
    c_{a\gamma\gamma}^{\rm eff} &= - \frac{\alpha_{\rm em}}{3 \pi} \cto \label{eq:ggctsimple} \\
    c_{a\gamma Z}^{\rm eff} &= \frac{2 \alpha_{\rm em} s_w}{3 \pi c_w} \cto \label{eq:gzctsimple} \\
    c_{aZZ}^{\rm eff} &= - \frac{\alpha_{\rm em} s_w^2}{3 \pi c_w^2} \cto \label{eq:zzctsimple} \\
    c_{aWW}^{\rm eff} &= 0 .\label{eq:wwctsimple}
\end{align}
These radiative corrections arise from triangle diagrams with the ALP and two gauge bosons connected via a top-quark loop, as shown in the left panel of Fig.~\ref{fig:alp-loops}.

\subsection{ALP couplings to fermions}

To assess the physical impact of ALP–fermion interactions at one-loop, it is essential to consider the renormalised, effective 
coupling \( c_f^{\text{eff}} \), which includes not only the tree-level contribution \( c_f \), but also radiative corrections. 
These arise from QED and QCD effects, as well as from loop-induced mixing with other fermionic and bosonic ALP couplings.

The complete expression for \( c_f^{\text{eff}} \), valid for an off-shell ALP and on-shell external fermions, is given by \cite{Bonilla:2021ufe}:
\begin{align}
\frac{c^{\text{eff}}_f}{f_a} &= \frac{c_f}{f_a} \left\{ 1 + \frac{\alpha_{\text{em}}}{2\pi} D_{c_f} + \frac{\alpha_s}{3\pi} D_{c_f}^g \right\} 
+ \frac{\alpha_{\text{em}}}{2\pi f_a} \left\{ c_{f'} D_{c_{f'}} + \sum_\psi c_\psi D_{c_\psi}^{\text{mix}} \right\} \nonumber \\
&\quad + \frac{\alpha_{\text{em}}}{2\pi} \left\{ g_{a\gamma\gamma} D_{\gamma\gamma} + g_{a\gamma Z} D_{\gamma Z} + g_{aZZ} D_{ZZ} + g_{aWW} D_{WW} \right\}
+ \frac{\alpha_s}{3\pi} \left\{ g_{agg} D_{gg} \right\}.
\label{eq:ceff}
\end{align}

The first term captures wave-function and vertex corrections proportional to the tree-level coefficient \( c_f \), while the second term accounts for mixing with other fermionic ALP couplings \( c_{f'} \). The third and fourth terms encode bosonic loop contributions arising from anomalous ALP–gauge interactions with photons, \( Z \), \( W \) bosons, and gluons. The loop functions \( D_i \) contain the full momentum dependence and finite parts of the one-loop integrals, and are given in \cite{Bonilla:2021ufe}.

Importantly, this expression shows that even when a tree-level coupling vanishes (\( c_f = 0 \)), a non-zero effective coupling can arise at low energies via radiative mixing with other sectors. 
The loop correction to the ALP-top coupling induced by $\cgo$, encoded in the function \( D_{gg} \) in Eq.~\eqref{eq:ceff}, is given by the expression:
\begin{equation}
\cteff = \cto + \frac{4\alpha_s}{3 \pi} \left[ 3 \log \left( \frac{\Lambda^2}{m_t^2} \right) - 4 - \frac{2 \pi^2}{3} - \frac{1}{2} \left( \log \left( \frac{m_t^2}{p^2} \right) + i \pi \right)^2 \right] \cgo
\label{eq:cteff}
\end{equation}
where \( \Lambda \) denotes the ultraviolet cutoff of the EFT. This correction depends logarithmically on the scale separation between the UV and the top mass, and also on the kinematic regime of the events used to constrain the coupling.

The corresponding loop diagram, in which the ALP couples to two gluons which in turn generate an ALP–top effective vertex, is shown on the right panel of Fig.~\ref{fig:alp-loops}.
\begin{figure}[t]
    \centering
    \includegraphics[width=\linewidth]{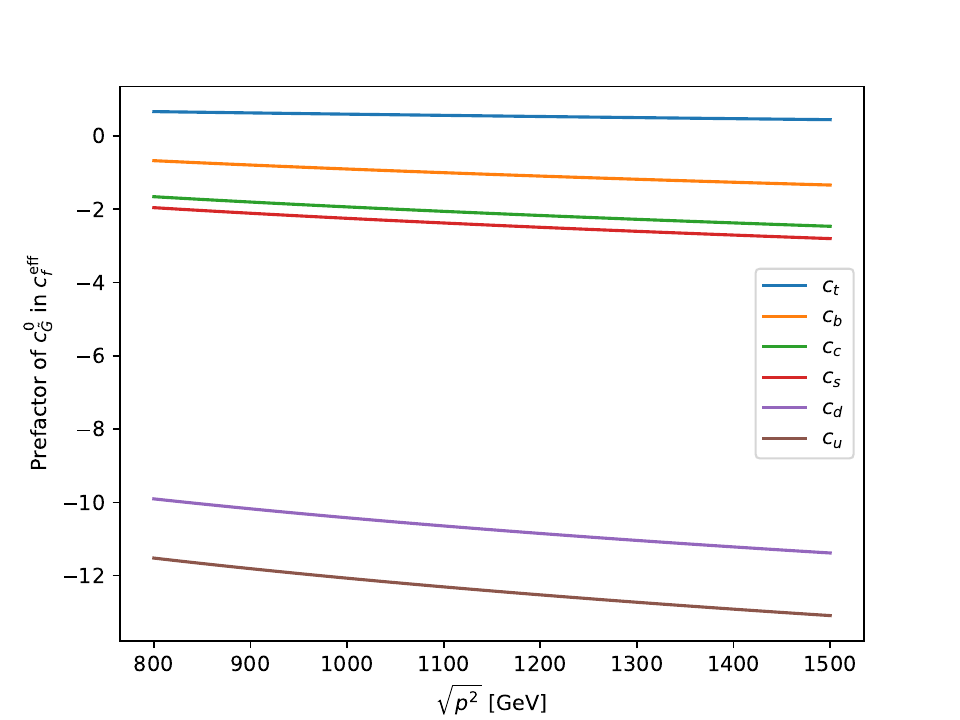}
    \caption{Coefficient multiplying \(\cgo\) for $\cteff$ in Eq.~\eqref{eq:cteff}, for $\cfeff$, $f = u,d,c,s$ in Eq.~\eqref{eq:ceff_highE} and for $\cbeff$ in Eq.~\eqref{eq:ceff_b_highE}, for \(\Lambda = 10\)~TeV as a function of the ALP energy $E = \sqrt{p^2}$.}
    \label{fig:pref_cgo}
\end{figure}
The blue line labeled $c_t$ in Fig.~\ref{fig:pref_cgo} shows the coefficient multiplying \(\cgo\) in Eq.~\eqref{eq:cteff} as a function of the ALP energy $\sqrt{p^2}$, fixing the cutoff to \(\Lambda = 10\)~TeV. In the relevant energy range, this prefactor varies between approximately 0.4 and 0.7. 

For instance, at $\sqrt{p^2} = 1$~TeV we obtain:
\begin{equation}
\cteff \simeq \cto + 0.6\, \cgo.
\label{eq:ctcgsimple}
\end{equation}

In the high-energy limit and assuming \(c_f^0 \ll \cto, \cgo\), ALP couplings to light fermions also receive significant loop-induced contributions. 
These include a term proportional to \(\cgo\), similar to Eq.~\eqref{eq:cteff}, but with the top quark mass replaced by the corresponding 
light fermion mass. Additionally, mixing effects—parameterized by \(D_{c_\psi}^{\text{mix}}\) in Eq.~\eqref{eq:ceff}—lead to terms proportional to \(\cto\).

For light quarks \(f = u, d, c, s\), the effective coupling takes the form~\cite{Bonilla:2021ufe}:
\begin{align}
\frac{c_f^{\text{eff}}}{f_a} &= \frac{4\alpha_s}{3 \pi f_a} \left[ 3 \log \left( \frac{\Lambda^2}{m_f^2} \right) - 4 - \frac{2 \pi^2}{3} - \frac{1}{2} \left( \log \left( \frac{m_f^2}{p^2} \right) + i \pi \right)^2 \right] \cgo \nonumber \\ 
&\quad - \frac{ \alpha_{\text{em}}}{2\pi f_a} \frac{3 T_{3,f} m_t^2}{2 s_w^2 M_W^2} \left[ \log\left( \frac{\Lambda^2}{p^2} \right) + 2 + i \pi \right] \cto.
\label{eq:ceff_highE}
\end{align}

For \(f = b\), an additional contribution from charged current interactions arises via \(D_{c_f'}\), leading to~\cite{Bonilla:2021ufe}:
\begin{align}
\frac{c_b^{\text{eff}}}{f_a} &= \frac{4\alpha_s}{3 \pi f_a} \left[ 3 \log \left( \frac{\Lambda^2}{m_b^2} \right) - 4 - \frac{2 \pi^2}{3} - \frac{1}{2} \left( \log \left( \frac{m_b^2}{p^2} \right) + i \pi \right)^2 \right] \cgo \nonumber \\ 
&\quad - \frac{ \alpha_{\text{em}}}{2\pi f_a} \frac{m_t^2}{8 s_w^2 M_W^2} \left[ -5\log\left( \frac{\Lambda^2}{p^2} \right)+\log\left( \frac{M_W^2}{p^2} \right) - \frac{17}{2} - 4 i\pi \right] \cto.
\label{eq:ceff_b_highE}
\end{align}

\begin{figure}[t]
    \centering
    \includegraphics[width=\linewidth]{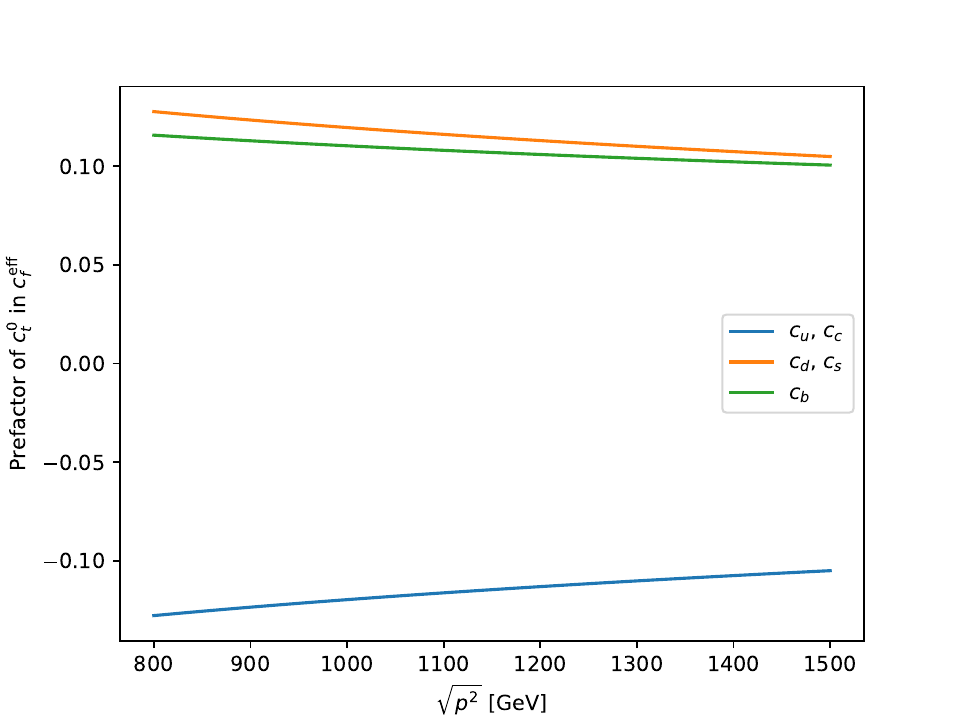}
    \caption{Coefficient multiplying \(\cto\) in Eq.~\eqref{eq:ceff_highE} and Eq.~\eqref{eq:ceff_b_highE}, for \(\Lambda = 10\)~TeV and various values of the ALP energy $E = \sqrt{p^2}$.}
    \label{fig:pref_cto}
\end{figure}

The prefactors of \(\cgo\) in Eqs.~\eqref{eq:cteff}, \eqref{eq:ceff_highE}, and \eqref{eq:ceff_b_highE} are shown in Fig.~\ref{fig:pref_cgo} as functions 
of \(p\) for fixed \(\Lambda = 10\)~TeV. The corresponding prefactors of \(\cto\) in Eqs.~\eqref{eq:ceff_highE} and \eqref{eq:ceff_b_highE} are shown in Fig.~\ref{fig:pref_cto}. 
The mixing contributions due to $\cto$ for the light quarks do not depend on the light quark mass but on their weak isospin, thus the prefactors for $c_u$ and $c_c$ coincide,  
as well as the prefactors for $c_d$ and $c_s$. The prefactor for $c_b$ differs because of the additional contribution induced by charged currents. 
Interestingly, the corrections are negative for up-type quarks and positive for down-type quarks.

Although effective ALP couplings to light fermions are induced, their magnitudes remain suppressed by small Yukawa couplings or electroweak mixing angles. 
In scenarios where \(\cgo = 0\), it has been shown in Ref.~\cite{Esser:2023fdo} that these couplings are sufficiently small that the ALP is effectively 
collider-stable, escaping the detector and yielding a missing transverse momentum signature. 
However, when \(\cgo \neq 0\), additional contributions may induce faster ALP decays into SM fermions. While decays to neutrinos retain the missing 
energy signature, heavier ALPs might decay to hadrons or charged leptons.

However, in many motivated models, the ALP also couples to exotic invisible states, such as dark sector particles, which may dominate the decay 
width and preserve the missing energy signature. For this reason, we assume in our study that ALPs produced in the final state—and their decay products—are invisible at the detector level.

Nevertheless, we separately evaluate constraints arising from off-shell ALP-mediated processes such as \(pp \to a \to \bar{t}t\) and \(pp \to a \to gg\), which 
do not rely on missing energy signatures.


%
\section{Model Building Setup}
\label{sec:models}
A scenario where leading-order couplings to tops and/or gluons appear at high energies naturally arises in models where the ALP is a composite pseudo-Nambu--Goldstone boson (pNGB), associated with the spontaneous breaking of a global symmetry in a strongly coupled sector. These models, inspired by composite Higgs frameworks or extra-dimensional theories, often incorporate partial compositeness, whereby SM fermions couple linearly to operators in the strong sector.

If the top quark is mostly elementary or neutral under the global symmetry responsible for the ALP, the tree-level coupling $\cto$ vanishes. However, a non-zero gluonic coupling $\cgo$ can still be present through anomaly-like interactions or loops of heavy coloured states. In this case, radiative corrections involving loops with gluons and a top, such as in the right panel of Fig.~\ref{fig:alp-loops},  generate an effective $\cteff$ at low energies, as shown in Eq.~\eqref{eq:ctcgsimple}.

Conversely, if the top quark is strongly mixed with composite fermions charged under the ALP symmetry, a tree-level coupling $\cto \neq 0$ is naturally induced. In this regime, $\cgo$ may be absent at tree level, either because the heavy coloured fermions that mediate it are integrated out or are neutral under the relevant symmetries. Nevertheless, loop corrections involving the top quark induce an effective coupling to gluons, modifying $\cgeff$ as shown in Eq.~\eqref{eq:cgctsimple}.

These two limiting cases illustrate how collider observables can help disentangle the ultraviolet structure of ALP interactions, shedding light on the symmetry assignments and compositeness of SM fields. We stress that the mapping between bare and effective couplings depends logarithmically on the UV cutoff. For instance, taking $\Lambda = 10$~TeV as in our plots, the correction is modest in absolute terms but can significantly affect small couplings. In particular, a positive $\cto$ can result in a negative $\cteff$ when the loop-induced shift is dominant.

\subsection{An Explicit Example: the \texorpdfstring{SO(6)/SO(5)}{} Composite Higgs Model}

To provide a concrete illustration of the possible UV origin of the effective couplings studied in this work, we consider an explicit model in which an ALP emerges from a well-motivated extension of the Standard Model. In particular, we focus on a class of composite Higgs models where both the Higgs boson and the ALP arise as pNGBs associated with the spontaneous breaking of a global symmetry in a strongly coupled sector. These scenarios not only naturally explain the lightness of the Higgs and ALP, but also generate the couplings of the ALP to Standard Model fields through well-defined mechanisms such as partial compositeness and anomaly matching. Among these, the next-to-minimal composite Higgs model based on the coset $SO(6)/SO(5)$ provides a minimal and predictive framework, which we now describe in more detail.

In the next-to-minimal composite Higgs model based on the coset $SO(6)/SO(5)$~\cite{Gripaios:2009pe, Contino:2011np, Ferretti:2013kya, Sanz:2015sua}, the Higgs doublet arises as a pNGB from the spontaneous breaking of a global $SO(6)$ symmetry. This coset structure also predicts a real singlet scalar $a$, which is CP-odd and thus a natural ALP candidate. The field \( a \) couples to SM fields through both anomaly-induced and loop-generated interactions. Of particular interest are its couplings to gluons and to top quarks, which arise from three main sources: an anomaly-induced Wess–Zumino–Witten (WZW) term stemming from a mixed \( SO(6)^2 \times SU(3)_C \) anomaly in the UV strong sector; loop diagrams involving the top quark, generated via partial compositeness; and loop contributions from heavy top partners, i.e. composite fermions charged under QCD and the ALP symmetry.

The effective Lagrangian for the couplings of $a$ to top quarks and gluons is:
\begin{align}
\mathcal{L}_{\text{eff}} \supset 
& - i\, \frac{c_t}{f_a}\, a\, \bar{t} \gamma_5 t \label{eq:eta-top-coupling} \\
& + \frac{\alpha_s}{8\pi f_a}\, a\, G^a_{\mu\nu} \tilde{G}^{a\,\mu\nu} \left( c_{\text{WZW}} + c_{\text{top}} + c_{\text{partners}} \right), \label{eq:eta-gluon-coupling}
\end{align}
where \(c_t\) parametrizes the pseudo-scalar Yukawa coupling to the top quark, typically of order \(m_t / f\) and determined by the mixing strength and group embedding of the top.

The gluon coupling receives three contributions:
\begin{align}
c_{\text{WZW}} &= \text{Anomaly coefficient from the UV strong sector}, \\
c_{\text{top}} &= \sum_f \kappa_f\, A_{1/2}^a(\tau_f) \quad \text{(top loops)}, \\
c_{\text{partners}} &= \sum_F \kappa_F\, A_{1/2}^a(\tau_F) \quad \text{(top partner loops)},
\end{align}
where $\tau_f = 4 m_f^2 / m_a^2$ and $A_{1/2}^a(\tau)$ is the standard pseudoscalar loop function, see App.~\ref{sec:appendix}, Eq.~(\ref{eq:A12}). In the limit \(\tau \gg 1\), i.e. when the fermion in the loop is much heavier than the ALP, the function simplifies to \(A_{1/2}^a(\tau) \to 2\).

The anomaly-induced contribution $c_{\text{WZW}}$ depends on the fermionic content of the strong sector. If the UV completion contains $N_\Psi$ Dirac fermions $\Psi$ in the fundamental of both $SO(6)$ and $SU(3)_C$, each with $a$-charge normalized to 1, then each contributes $c_{\text{WZW}}^{(i)} = 2 T(R) Q_a = 2 \cdot \frac{1}{2} \cdot 1 = 1,$
yielding a total anomaly coefficient
\begin{equation}
c_{\text{WZW}} = \sum_{i=1}^{N_\Psi} c_{\text{WZW}}^{(i)} = N_\Psi.
\end{equation}
Typical values in composite Higgs UV completions are \(c_{\text{WZW}} \sim \mathcal{O}(1\text{--}10)\)~\cite{Ferretti:2013kya}.

Loop-induced contributions $c_{\text{top}}$ and $c_{\text{partners}}$ depend on the details of the top embedding and top partner spectrum. For example, if the top is embedded in the $\mathbf{6}$ of $SO(6)$, $c_t$ can be non-zero at leading order, contributing both to direct top couplings and to gluon couplings via loops. Top partner contributions can be important if the partners are not too heavy and couple significantly to the ALP. The contribution from top partners, encoded in \( c_{\text{partners}} \), depends on the spectrum and couplings of the heavy composite fermions charged under QCD and the global symmetry associated with the ALP. In models where the top partners have masses \( M_T \sim 1.5\text{--}2.5\ \mathrm{TeV} \) and sizable pseudo-scalar couplings to the ALP, the loop-induced contribution can be significant. Assuming a typical coupling \( \kappa_T \sim \mathcal{O}(1) \) and \( m_a \ll M_T \), the loop function approaches \( A_{1/2}^a(\tau_T) \to 2 \), yielding
\[
c_{\text{partners}} \sim 2 \kappa_T \sim \mathcal{O}(1\text{--}2).
\]
These contributions can partially cancel or enhance the top loop or anomaly terms in Eq.~\eqref{eq:eta-gluon-coupling}, and are thus an important ingredient in determining the low-energy phenomenology of ALP couplings to gluons.

In summary, typical values in UV-complete composite Higgs models yield \( \cto \sim 1 \) and \( \cgo \sim 1\text{--}10 \), depending on the strength of top compositeness and the number of colored fermions generating the mixed anomaly.

\section{LHC Constraints on ALP Couplings}
\label{sec-results}

In this section, we present the phenomenological implications of the ALP couplings introduced in the previous sections.
We explore the sensitivity of current LHC measurements to the effective ALP interactions with gluons and top quarks, taking into account the loop-induced modifications discussed in Section~\ref{sec:theory}.
By confronting our effective theory with experimental data using the \contur framework, we derive bounds on the relevant Wilson coefficients, and identify regions of the parameter space where new physics effects could still be present.
In particular, we focus on the interplay between the effective couplings \( \cgeff \) and \( \cteff \), defined in Eqs.\ (\ref{eq:cgctsimple}) and (\ref{eq:cteff}), and the role of radiative corrections in shaping the phenomenological constraints.

\subsection{Effective couplings}

\begin{figure}[tbph]
    \centering
    
    \begin{subfigure}[t]{\linewidth}
        \includegraphics[width=\linewidth]{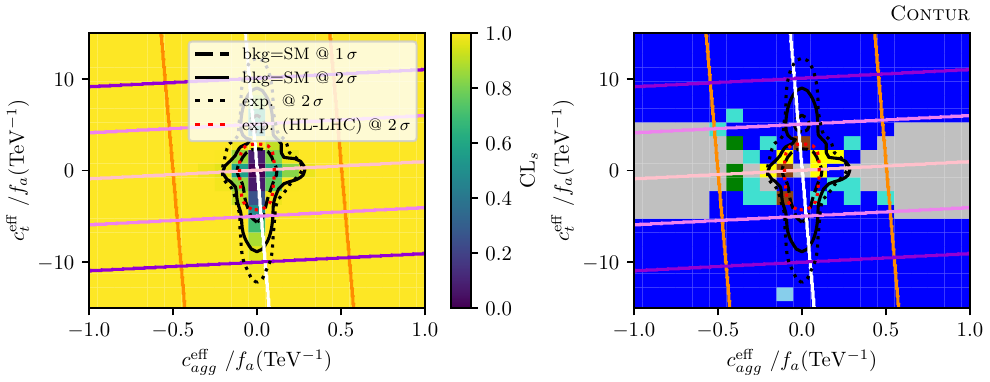}
    \end{subfigure}
        
    \begin{subfigure}[t]{\linewidth}
        \includegraphics[width=\linewidth]{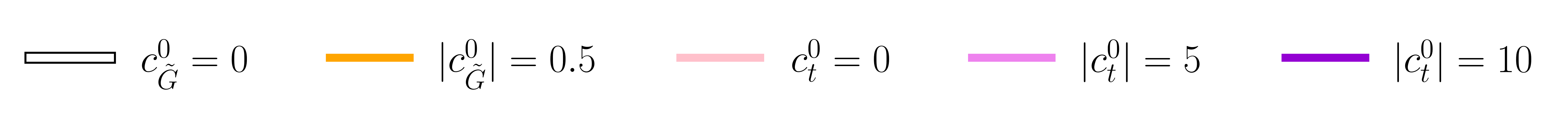}
    \end{subfigure}
    
    \begin{subfigure}[t]{\linewidth}
        \DeclareRobustCommand{\swatch}[1]{\tikz[baseline=-0.6ex]\node[fill=#1,shape=rectangle,draw=black,thick,minimum width=5mm,rounded corners=0.5pt](){};}

\definecolor{blue}{HTML}{0000FF}
\definecolor{yellow}{HTML}{FFFF00}
\definecolor{black}{HTML}{000000}
\definecolor{saddlebrown}{HTML}{8B4513}
\definecolor{green}{HTML}{008000}
\definecolor{skyblue}{HTML}{87CEEB}
\definecolor{turquoise}{HTML}{40E0D0}
\definecolor{silver}{HTML}{C0C0C0}

    \begin{tabular}{lll}
        \swatch{blue}~$\ell$+\MET{}+jet \cite{CMS:2021vhb,ATLAS:2017cez,ATLAS:2019hxz,ATLAS:2015mip,ATLAS:2017luz,ATLAS:2022xfj,ATLAS:2017irc,CMS:2016oae,CMS:2018tdx,CMS:2018htd,ATLAS:2015lsn} & 
        \swatch{yellow}~$\gamma$ \cite{ATLAS:2017xqp,ATLAS:2024vqf,ATLAS:2019iaa} & 
        \swatch{saddlebrown}~hadronic $t\bar{t}$ \cite{ATLAS:2018orx,ATLAS:2020ccu,CMS:2019fak,ATLAS:2022mlu,CMS:2019eih} \\
        \swatch{green}~\MET{}+jet \cite{ATLAS:2017txd,ATLAS:2024vqf} & 
        \swatch{skyblue}~$\ell_1\ell_2$+\MET{} \cite{CMS:2022woe,CMS:2020mxy} & 
        \swatch{turquoise}~$\ell_1\ell_2$+\MET{}+jet \cite{ATLAS:2023gsl,ATLAS:2019hau,ATLAS:2019ebv} \\
        \swatch{silver}~jets \cite{ATLAS:2015xtc,ATLAS:2017ble} &
        \swatch{orange}~$\ell^+\ell^-$+jet \cite{ATLAS:2017nei,ATLAS:2019ebv,ATLAS:2024tnr,ATLAS:2020juj}
    \end{tabular}

    \end{subfigure}
    
    \caption{Exclusions in the \cgeff--\cteff parameter space using \contur. 
      (Top Left) Grid of exclusions and interpolated contours. (Top Right) Most exclusionary analysis per point. (Bottom) Legend of dominant analyses.
      The overlaid grid lines show constant bare coupling values.}
    \label{fig:contur_top1}
\end{figure}

Figure~\ref{fig:contur_top1} shows the constraints in the plane of the effective ALP couplings to gluons (\cgeff) and to top
quarks (\cteff), obtained using the \contur framework.
The analysis assumes a fixed ALP mass of $m_a = 1~\mathrm{GeV}$, although masses below {\cal O}(100) GeV would lead to the same
results, under the assumption that the ALP either remains collider-stable or decays to invisible particles.

The left panel presents the exclusion regions derived from current LHC measurements, with $1\sigma$ and $2\sigma$
confidence levels shown for a SM background hypothesis. Also shown is the expected $2\sigma$ limit
(obtained if the measurement had agreed exactly with the SM), as well as an estimate of the similarly expected $2\sigma$
exclusion for the High-Luminosity LHC (HL-LHC)
\footnote{This is based on simple luminosity scaling, which is likely to be pessimistic, see for example~\cite{Belvedere:2024wzg}.}. 

The right panel identifies the most sensitive signal at each point in parameter space, revealing the complementarity of different
final states.
It can be seen that $\ell + \ETmiss + \text{jet}$, which is a final state produced by semileptonic \ttbar events,
dominates most of the plane where $\cteff \gtrapprox 1$, with the jet measurements dominating for most of the region of smaller \cteff.
Other measurements also contribute, as will be discussed further below. 
This demonstrates the ability of LHC measurements to probe different signatures across the ALP parameter space,
with current data already placing significant constraints and HL-LHC prospects extending sensitivity further.

For this scan, the effective couplings \cteff and \cgeff are treated as free parameters, which are related to the bare parameters
\cto and \cgo by Eqs.~(\ref{eq:cgctsimple}) and (\ref{eq:ctcgsimple}). 
The relations in Eqs.~(\ref{eq:ggctsimple})-(\ref{eq:zzctsimple}) also imply non-zero couplings to photons and $Z$-bosons,
which are calculated in this case from \cgeff and \cteff, via the bare coupling relations. The mapping between
the (\cteff, \cgeff) plane and the (\cto, \cgo) planes is indicated by the superimposed coloured grid lines for $\Lambda = 10$~TeV.

\subsection{Bare couplings}

\begin{figure}[tbph]
   \centering
   
   \includegraphics[width=\linewidth]{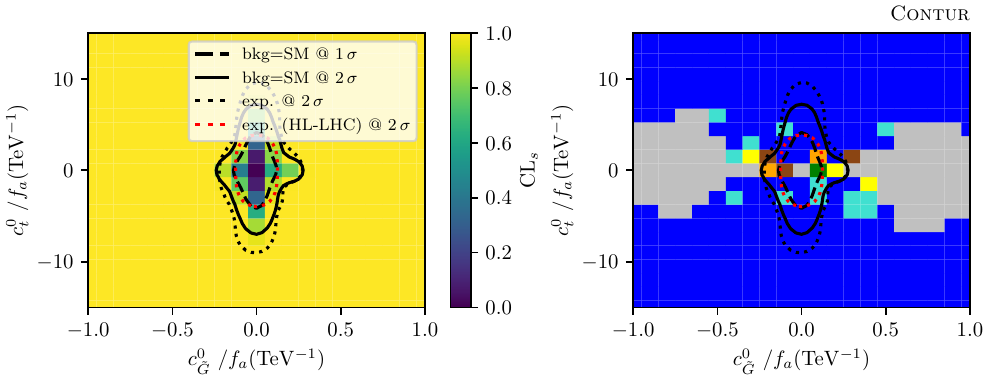}
   
   \caption{Exclusions in the \cgo - \cto parameter space using \contur, assuming $\Lambda=10$~TeV.
     The legend for the dominant analysis at each point is as in Fig.~\ref{fig:contur_top1}.}
   \label{fig:contur_top2}
\end{figure}

In Fig.~\ref{fig:contur_top2}, \cto and \cgo are treated as free parameters, with the effective couplings calculated
according to Eqs.~(\ref{eq:cgctsimple})-(\ref{eq:wwctsimple}) and (\ref{eq:ctcgsimple}). This is a slightly more model-dependent approach than
constraining the effective parameters, but under our assumptions that \cto and \cgo are the only non-zero tree-level couplings, it is more
natural to discuss in these terms.

When evaluting exclusions, \contur groups measurements into ``pools'', based upon
beam energy, experiment (generally ATLAS or CMS) and final state, such that there are no statistical
correlations (that is, no shared events) between different pools~\cite{Butterworth:2016sqg}.
For a given parameter point, only the most exclusionary measurement from each pool is then used, with the others discarded
to avoid double-counting. The independent exclusions from each pool are then combined to give the overall exclusion.
To look in more detail at where the sensitivity to the ALP comes from, we can look at the exclusion from
relevant example pools individually.

\begin{figure}[tbp]
    \centering
    
    \begin{minipage}{\textwidth}
        \begin{subfigure}[t]{0.45\textwidth}
            \centering
            \includegraphics[width=\textwidth,height=5cm,keepaspectratio]{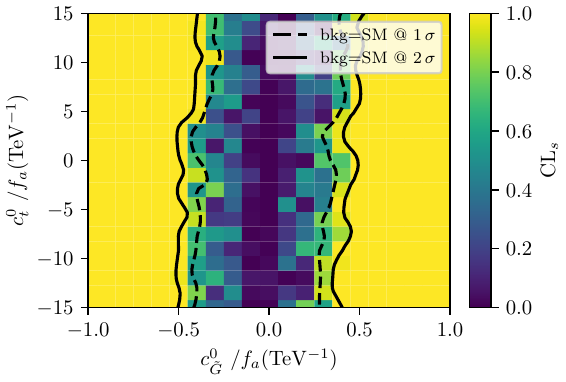}
            \caption{\label{fig:dijetexc}}
        \end{subfigure}
        \hfill
        \begin{subfigure}[t]{0.45\textwidth}
            \centering
            \includegraphics[width=\textwidth,height=5cm,keepaspectratio]{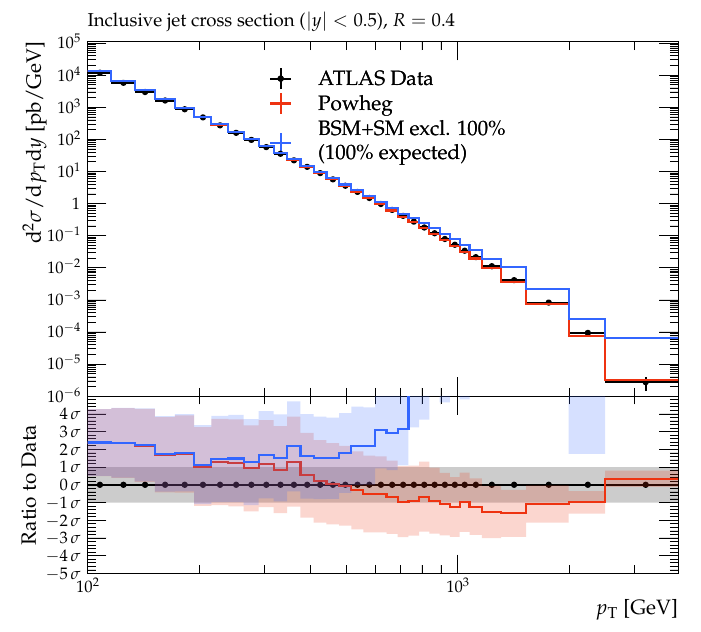}
            \caption{\label{fig:dijethist}}
        \end{subfigure}
        \caption{
          (a) Exclusions from ATLAS measurements in the dijet final state~\cite{ATLAS:2015xtc,ATLAS:2017ble}.
          (b) Exclusionary histogram from Ref.~\cite{ATLAS:2017ble}, where \cgo=0.7, \cto=0\label{fig:dijet}.
          The lower inset shows the difference between data, SM~\cite{Altakach:2020ugg} and SM+BSM, in terms of the standard deviation of the data.}
    \end{minipage}
\end{figure}

In Fig.~\ref{fig:dijetexc} we show the exclusion coming from the pool containing ATLAS 13~TeV dijet measurements.
In this case, despite the contribution to \cgeff generated by a non-zero \cto, the exclusion does not strongly depend on \cto;
the dependence is manifest as a small gradient in the exclusion bands, with the exclusion being slightly stronger when \cto and \cgo
have opposite signs,
since then a non-zero \cto will have a constructive impact on \cgeff according to Eq.~(\ref{eq:cgctsimple}),
leading to to an increased cross section for ALP-mediated dijet production.
The dijet measurements exclude values of $|\cgo| \gtrapprox 0.5$ for $\cto=0$. Fig.~\ref{fig:dijethist}
shows an example of one of the measurements providing exclusion at an example parameter point ($\cgo = 0.7, \cto = 0$).
The deviation from the SM that would have been seen if these values were realised in nature is clearly seen at high jet
transverse momentum ($\pt$). 

\begin{figure}[tbp]    
    \begin{minipage}{\textwidth}
        \begin{subfigure}[t]{0.45\textwidth}
            \centering
            \includegraphics[width=\textwidth,height=5cm,keepaspectratio]{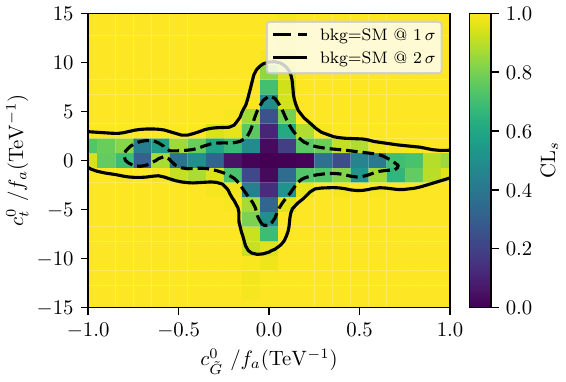}
            \caption{\label{fig:cmstlexc}}
        \end{subfigure}
        \hfill
        \begin{subfigure}[t]{0.45\textwidth}
            \centering
            \includegraphics[width=\textwidth,height=5cm,keepaspectratio]{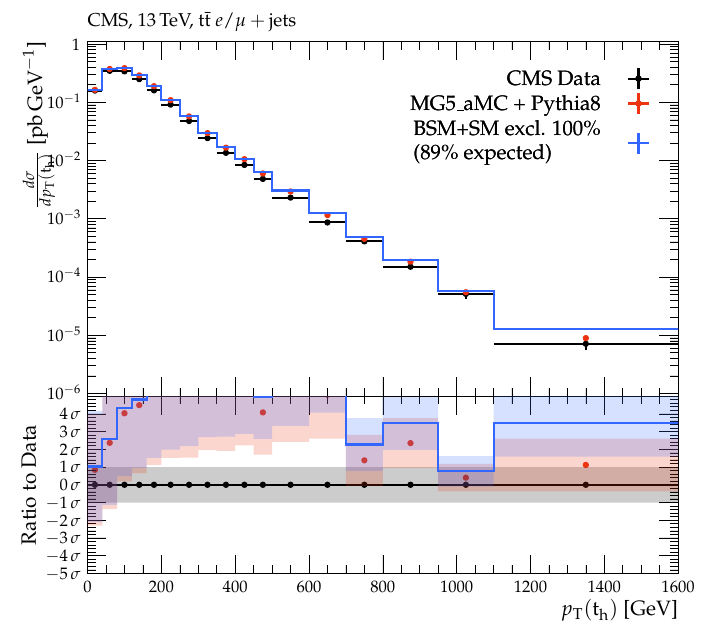}
            \caption{\label{fig:cmstlhist}}
        \end{subfigure}
        \caption{(a) Exclusions from CMS measurements in final states with a lepton, jets and missing energy~\cite{CMS:2021vhb,CMS:2016oae,CMS:2018tdx,CMS:2018htd}. (b) Exclusionary histogram from Ref.~\cite{CMS:2021vhb}, where \cgo=0, \cto=12.0\label{fig:cmstl}.
        The lower inset shows the difference between data, SM~\cite{Alwall:2014hca,Sjostrand:2014zea} and SM+BSM, in terms of the standard deviation of the data.}
    \end{minipage}
\end{figure}

In Fig.~\ref{fig:cmstlexc} we show the exclusion coming from the pool containing CMS 13~TeV measurements
of the $\ell$+\MET{}+jet final state. This final state includes $W$+jet processes and \ttbar processes with
one top decaying hadronically and the other leptonically. This is the most powerful channel for excluding
ALP processes when $|\cto| \gtrapprox 2$ for most values of \cgo, as can be seen in Fig.~\ref{fig:contur_top2}.
Even when $\cto = 0$, this channel provides some sensitivity if \cgo is large enough, due to the
induced \cteff.
Fig.~\ref{fig:cmstlhist} shows an example of one of the measurements providing exclusion at an example parameter point
($\cgo = 0, \cto = 12$) - in this case the hadronic top \pt. The SM does not describe this distribution
particuarly well (the $p$ value is 0.09 if the theory uncertainties are assumed uncorrelated) but the ALP contribution
would take it event further away from the data, to the point where $p < 0.05$. 

\begin{figure}[tbp]    
    
    \begin{minipage}{\textwidth}
        \begin{subfigure}[t]{0.45\textwidth}
            \centering
            \includegraphics[width=\textwidth,height=5cm,keepaspectratio]{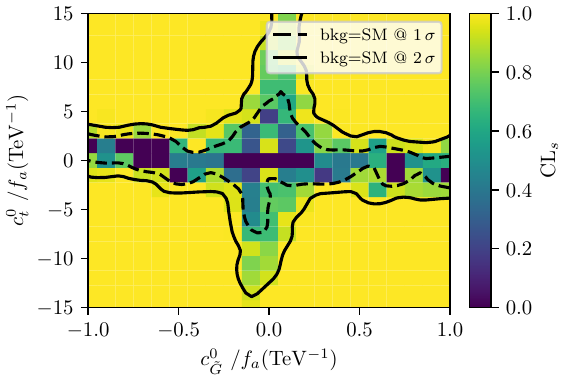}
            \caption{}
        \end{subfigure}
        \hfill
        \begin{subfigure}[t]{0.45\textwidth}
            \centering
            \includegraphics[width=\textwidth,height=5cm,keepaspectratio]{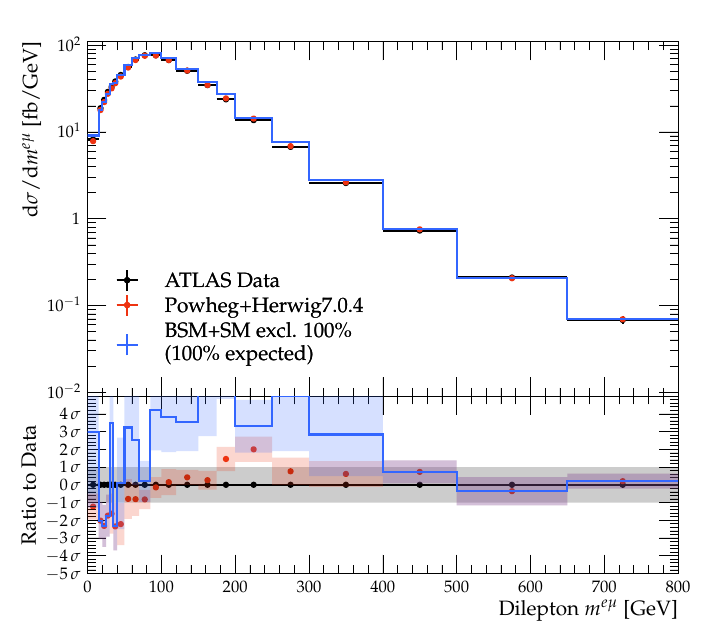}
            \caption{}\label{fig:8b}
        \end{subfigure}
        \caption{
          (a) Exclusions from ATLAS $t\bar t$ measurements in the dileptonic final state~\cite{ATLAS:2019ebv,ATLAS:2023gsl,ATLAS:2019hau}.
          (b) Exclusionary histogram from Ref.~\cite{ATLAS:2023gsl}, where \cgo=0.5, \cto=4.5.\label{fig:ttll}
        The lower inset shows the difference between data, SM~\cite{Alioli:2010xd,Bellm:2015jjp} and SM+BSM, in terms of the standard deviation of the data.}
    \end{minipage}
    
    \begin{minipage}{\textwidth}
        \begin{subfigure}[t]{0.45\textwidth}
            \centering
            \includegraphics[width=\textwidth,height=5cm,keepaspectratio]{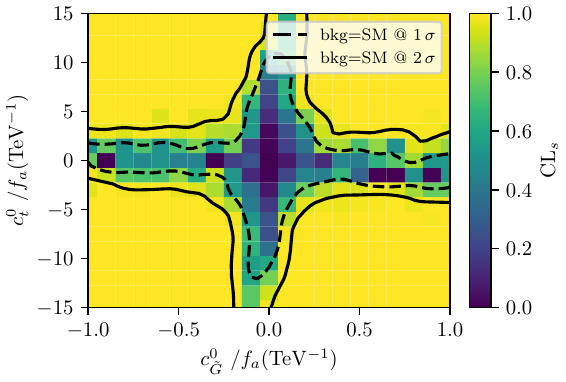}
            \caption{}
        \end{subfigure}
        \hfill
        \begin{subfigure}[t]{0.45\textwidth}
            \centering
            \includegraphics[width=\textwidth,height=5cm,keepaspectratio]{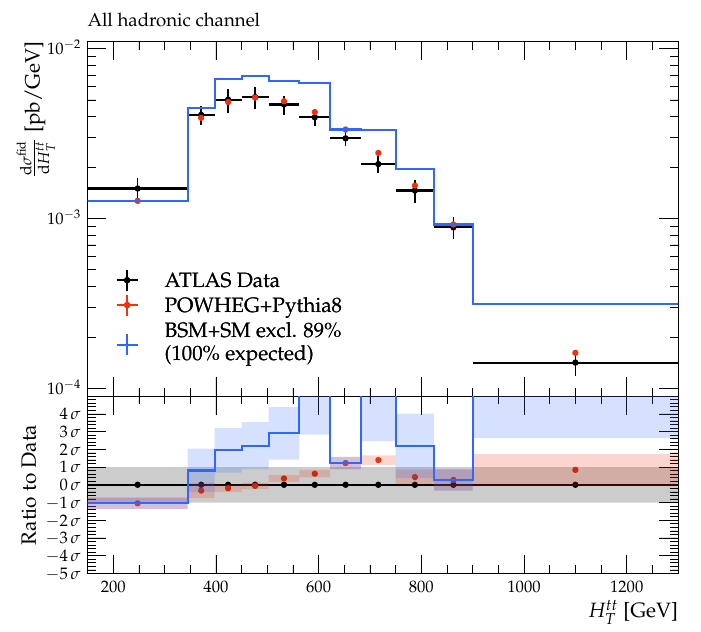}
            \caption{} \label{fig:9b}
        \end{subfigure}
        \caption{
          (a) Exclusions from ATLAS $t\bar t$ measurements in the all-hadronic final state~\cite{ATLAS:2018orx,ATLAS:2020ccu,ATLAS:2022mlu}.
          (b) Exclusionary histogram from Ref.~\cite{ATLAS:2020ccu}, where \cgo=0.6, \cto=6.0.\label{fig:tthh}
        The lower inset shows the difference between data, SM~\cite{Frixione:2007vw,Sjostrand:2014zea} and SM+BSM, in terms of the standard deviation of the data.}
    \end{minipage}
\end{figure}

The other \ttbar signatures -- fully leptonic and fully hadronic -- also give strong exclusions, as shown in
Fig.~\ref{fig:ttll} and Fig.~\ref{fig:tthh} respectively, along with examples of the measurements provide the
sensitivity.

\begin{figure}[tbp]
    \centering
    
    \begin{minipage}{\textwidth}
        \begin{subfigure}[t]{0.45\textwidth}
            \centering
            \includegraphics[width=\textwidth,height=5cm,keepaspectratio]{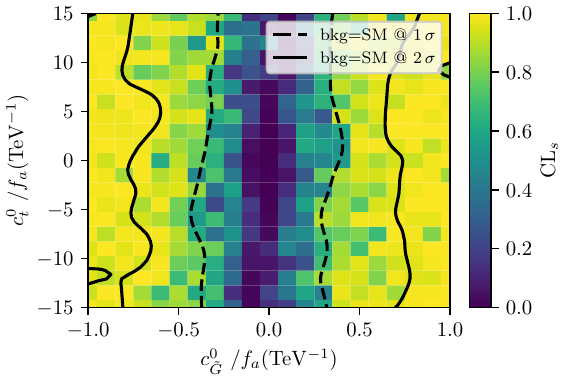}
            \caption{\label{fig:gamexc}}
        \end{subfigure}
        \hfill
        \begin{subfigure}[t]{0.45\textwidth}
            \centering
            \includegraphics[width=\textwidth,height=5cm,keepaspectratio]{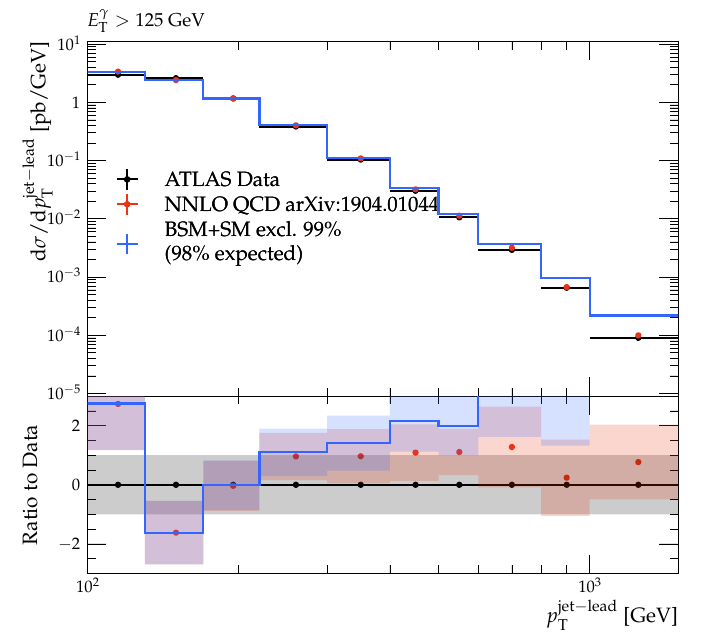}
            \caption{\label{fig:gamhist}}
        \end{subfigure}
        \caption{
          (a) Exclusions from ATLAS measurements in final states with an isolated photon and jets~\cite{ATLAS:2017xqp,ATLAS:2024vqf,ATLAS:2019iaa}.
          (b) Exclusionary histogram from Ref.~\cite{ATLAS:2017xqp}, where \cgo=0.6, \cto=1.5.
        The lower inset shows the difference between data, SM~\cite{Chen:2019zmr} and SM+BSM, in terms of the standard deviation of the data.}
    \end{minipage}
\end{figure}

\begin{figure}[tbp]
    \centering
    
    \begin{minipage}{\textwidth}
        \begin{subfigure}[t]{0.45\textwidth}
            \centering
            \includegraphics[width=\textwidth,height=5cm,keepaspectratio]{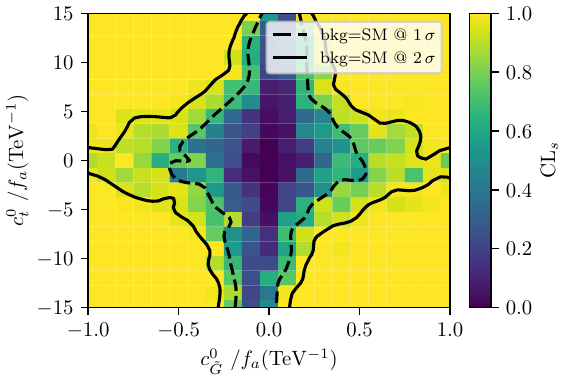}
            \caption{\label{fig:zexc}}
        \end{subfigure}
        \hfill
        \begin{subfigure}[t]{0.45\textwidth}
            \centering
            \includegraphics[width=\textwidth,height=5cm,keepaspectratio]{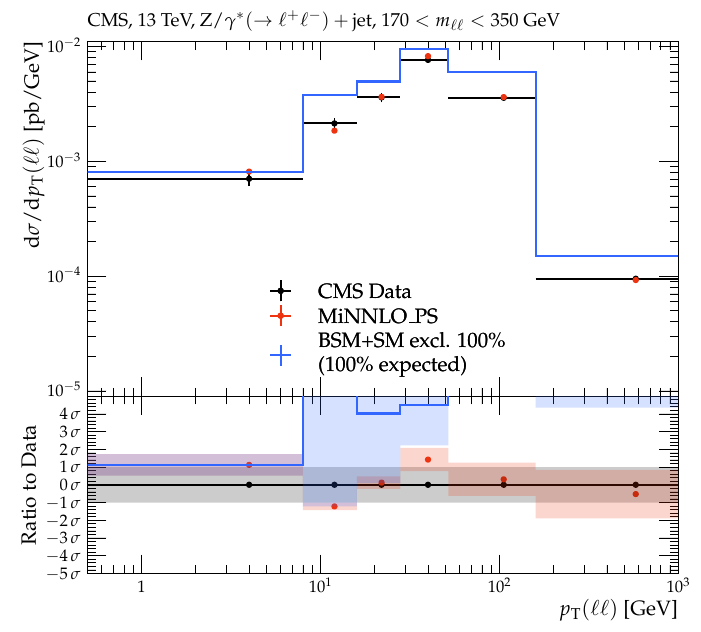}
            \caption{\label{fig:zhist}}
        \end{subfigure}
        \caption{
          (a) Exclusions from CMS measurements in the dilepton+jets final state~\cite{CMS:2018mdf,CMS:2019raw,CMS:2021wfx,CMS:2022ubq}.
          (b) Exclusionary histogram from Ref.~\cite{CMS:2022ubq}, where \cgo=0.6, \cto=7.5.
        The lower inset shows the difference between data, SM~\cite{Monni:2019whf,Monni:2020nks} and SM+BSM, in terms of the standard deviation of the data.}
    \end{minipage}
\end{figure}

Finally, and less expectedly, there are contributions from vector boson associated production with jets. These
arise from the radiative production of an ALP from a gluon line in $W,Z$ or $\gamma$ plus jets production. Due to the
overlapping final states, $W$+jets contributes to the pool already shown in Fig.~\ref{fig:cmstl}. Although the QCD background for
$\gamma$+jets is large, this final state does provide some sensitivity, as shown in Fig~\ref{fig:gamexc}. The signal cross section for this process is smaller than other signal processes considered, leading some noise in the exclusions from limited statistics. The effect of the ALP in this process can be seen in Fig.~\ref{fig:gamhist}, where ALP radiation from an initial state gluon leads to a hardening of the jet \pt spectrum, which is less consistent with data than the SM prediction from Ref.~\cite{Chen:2019zmr}. More sensitivity comes from $Z$+jets, as shown in
Fig.~\ref{fig:zexc} for the example of CMS 13~TeV data. As can be seen in Fig.~\ref{fig:zhist}, this comes from
an enhancement in the (precisely measured) $Z$ production cross section at high $Z$ \pt, which is inconsistent with the data.

\subsection{Impact of virtual ALP exchange alone}

\begin{figure}[tbp]
   \centering
   
   \includegraphics[width=\linewidth]{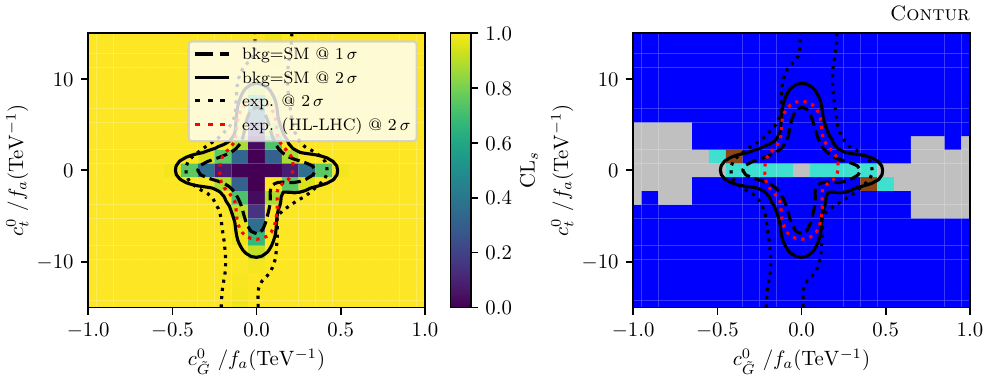}
   
   \caption{Exclusions in the \cgo - \cto parameter space using \contur, from $t\bar{t}$ and dijet diagrams mediated by a virtual ALP, i.e.\ we exclude signal processes with ALPs in the final state. Again we assume  $\Lambda=10$~TeV, and the legend of most exclusionary analyses is as in Fig.~\ref{fig:contur_top1}.}
   \label{fig:contur_top3}
\end{figure}

Since much of the above sensitivity comes from processes in which the ALP is in the final state, the results rely on the assumption
that the ALP, or its decay products, are invisible, and are thus manifest as $\ETmiss$.
Should the ALP in fact decay in the detector, other signatures, possibly including displaced vertices, would come into play.
This could increase or reduce the sensitivity, in a rather model-dependent fashion.
To explore the maximal adverse impact of this model dependence on the sensitivity, in
Fig.~\ref{fig:contur_top3} we show the exclusion in the \cto-\cgo plane when ALP production diagrams are switched off, and thus the 
sensitivity comes only from processes in which a virtual ALP is a mediator in the diagram.
This represents the weakest possible constraint, based on the assumption that ALP production itself provides no contribution.

Comparing to Fig.~\ref{fig:contur_top2}, we see that the limits are weakened, particularly along the $\cto = 0$ and $\cgo = 0$ axes,
where the $|\cto|$ limit moves from around 7 to around 10, and the $|\cgo|$ limit moves from around 0.3 to around 0.5.

\section{ALP EFT Validity at high-momentum transfer}
\label{sec:validity}
In this study, we employ an effective field theory framework to describe the interactions of the axion-like particle with Standard Model fields. While this approach is robust at energies well below the scale of new physics, its applicability at high momentum transfer requires careful assessment, particularly since LHC processes probe the multi-TeV regime.

Following the discussion in Ref.~\cite{Esser:2023fdo}, we examine the validity of the EFT in the context of our analysis. The limits shown in Figs.~\ref{fig:contur_top2} and \ref{fig:contur_top3} were obtained without imposing an explicit EFT validity cut, under the assumption that the expansion remains valid across the full kinematic range relevant to the LHC observables considered. This assumption must be tested, especially in regions of parameter space where the momentum transfer approaches or exceeds the EFT cutoff scale. 

From the distributions in Figs.~\ref{fig:dijethist}, \ref{fig:cmstlhist}, \ref{fig:8b}, \ref{fig:9b}, \ref{fig:gamhist} and \ref{fig:zhist}, one sees that the kinematic region where the ALP signal stands out above the SM background is typically in the few-hundred-GeV range.  To ensure the validity of the EFT expansion in powers of $E/f_a$, we therefore require that the characteristic energy scale $\sqrt{\hat{s}}$ of each event satisfies $\sqrt{\hat{s}} < f_a$ for the values of $c_i$ under consideration. In addition, the expansion assumes that the relevant momentum transfer is larger than the mass of the particles integrated out; in our case this means $p \gg m_t$. This condition is well satisfied in practice, since the most constraining regions of our analysis arise from events with $p \gtrsim 200~\text{GeV}$.

A more careful assessment involves comparing the characteristic energy scale of the events, typically the partonic center-of-mass energy $\sqrt{\hat s}$, to the cutoff scale, which we identify with the decay constant $f_a$ associated with the ALP. In particular, in Figs.~\ref{fig:dijethist}, \ref{fig:cmstlhist}, \ref{fig:8b}, \ref{fig:9b}, \ref{fig:gamhist} and~\ref{fig:zhist}, one sees that the kinematic regions where the ALP signal manifests most prominently above the SM background lie in the few hundred~GeV range. 

Therefore, for a given value of the dimensionless coupling $ c_i$, we must verify that the condition $\sqrt{\hat s} < f_a$ is satisfied across the relevant phase space. This ensures the convergence of the EFT expansion in powers of $E/f_a$ and justifies the interpretation of our bounds within a consistent low-energy effective theory.

In Figs.~\ref{fig:contur_top2} and~\ref{fig:contur_top3}, we present the 2$\sigma$ confidence level  exclusion limits in the plane of effective couplings \( (\cto/f_a,\, \cgo/f_a) \), derived under the assumption that the effective field theory (EFT) remains valid across the entire kinematic range. In particular, no cut has been applied on the partonic center-of-mass energy \( \sqrt{\hat{s}} \). The thick black contour delineates the excluded region. In the most constraining case, shown in Fig.~\ref{fig:contur_top2}, the limits on the bare top and gluon couplings at 2$\sigma$ correspond to \( f_a/\cto \gtrsim 200~\mathrm{GeV} \) and \( f_a/\cgo \gtrsim 5~\mathrm{TeV} \), respectively. Assuming \( f_a \gtrsim 1~\mathrm{TeV} \), these bounds imply that order-one values of \( \cto \) and \( \cgo \) remain compatible with current constraints. In particular, the limits obtained are not so weak as to demand nonperturbative values of the effective couplings for \( f_a \) in the TeV range, where ``nonperturbative'' refers to magnitudes exceeding \(\mathcal{O}(4\pi)\). This ensures that the EFT expansion remains under control in the relevant parameter space. Moreover, the resulting constraints are consistent with expectations from UV completions discussed in Sec.~\ref{sec:models}, where order-one  couplings naturally arise.

This shows that current Run~2 LHC data are already probing well-motivated UV scenarios, with sensitivity expected to improve further in Run~3 and HL-LHC runs.

\section{Conclusions}
\label{sec:concls}

In this work, we have studied the collider phenomenology of axion-like particles (ALPs) that couple dominantly to third-generation quarks and gluons, using an effective field theory (EFT) framework. We have focused on the loop-induced effects that generate couplings to gauge bosons and light fermions, even when tree-level interactions are absent. These effects were incorporated analytically, and their impact evaluated through reinterpretation of existing LHC measurements using the \contur framework.

We provided a careful treatment of the effective couplings, highlighting how top loops generate radiative corrections to ALP interactions with electroweak gauge bosons, and how gluon couplings induce corrections to the top and light fermions. These corrections can significantly alter the physical predictions of ALP couplings, especially in regimes where the bare couplings are small or vanishing. We illustrated these effects through both analytical expressions and representative values of momentum-dependent prefactors.

We explored  UV scenarios where the ALP couples only to gluons at high scale, and another where it couples dominantly to tops, or both. We showed how the low-energy effective interactions differ in each case and how collider observables can be used to discriminate among them. An explicit realisation was discussed within the context of the $SO(6)/SO(5)$ composite Higgs model, where both direct anomalous couplings and loop-induced effects contribute to the ALP interactions.

Using \contur, we derived limits in the effective coupling plane $c_t/f_a$ vs.\ $c_{\tilde{G}}/f_a$, based on a wide range of LHC measurements. Our results show that, even in the absence of dedicated ALP production searches, significant regions of parameter space can be excluded from precision SM measurements. In particular, the $2\sigma$ exclusion contours in Figs.~\ref{fig:contur_top2} and \ref{fig:contur_top3} demonstrate the power of this approach, especially in regions where the ALP is effectively invisible.

We also assessed the validity of the EFT approach by comparing the typical momentum transfer $\sqrt{\hat s}$ in the relevant signal regions to the effective cutoff scale $f_a$. We showed that while the EFT remains valid for a broad region of the parameter space probed, caution is needed when interpreting limits at high energy or large coupling, where the condition $\sqrt{\hat s} < f_a$ may be violated.

Looking ahead, this framework can be extended in several directions. One natural step is to consider the inclusion of couplings to third-generation leptons, such as the $\tau$, which can lead to complementary signatures at colliders. Another is to study the potential for displaced vertex signatures arising from long-lived ALPs, particularly in scenarios with suppressed decay widths. Finally, combining loop-aware EFT reinterpretations with direct ALP production searches — both visible and invisible — will offer a more complete picture of ALP phenomenology at the LHC and beyond.

\section*{Acknowledgements}

We would like to thank Annie Williams for preliminary studies and discussions.
The work of V.S. is supported by the Spanish grants  PID2023-148162NB-C21, CNS-2022-135688, and 
CEX2023-001292-S.
The work of M.U. is supported by the European Research
Council under the European Union’s Horizon 2020 research and innovation Programme (grant agreement
n.950246) and in part by STFC consolidated grant ST/X000664/1. F.E. is supported from Charles University through project PRIMUS/24/SCI/013.
J.C.E. is supported by the STFC UCL Centre for Doctoral Training in Data Intensive Science, grant ST/W00674X/1,
including departmental and industry contributions.
J.M.B. is supported in part by STFC consolidated grant ST/W00058X/1.

\appendix
\section{Loop functions}
\label{sec:appendix}
Here we list the loop functions used in Sec.\ \ref{sec:aVV}, following the conventions in \cite{Bonilla:2021ufe}.
The function $B_1$, that enters the effective couplings $\cgeff$, Eq.\ (\ref{eq:cagg_1L}), and $c_{a\gamma\gamma}^{\rm eff}$, Eq.\ (\ref{eq:cagammagamma_1L}), is given by \cite{Bauer:2017ris}:
\begin{flalign}
    B_1 (\tau) = 1 - \tau f^2 (\tau) \,, \qquad \mbox{with } f (\tau) =  \begin{cases} \arcsin \frac{1}{\sqrt{\tau}} & \mbox{for } \tau \ge 1 \\ \frac{\pi}{2} + \frac{i}{2} \ln{\frac{1 + \sqrt{1 - \tau}}{1- \sqrt{1 - \tau}}} & \mbox{for } \tau < 1 \end{cases}\,
\label{eq:B1}
\end{flalign}
The functions $A^{\text{t}}$, $B^{\text{t}}$ and $C^{\text{t}}$ in $c_{a\gamma Z}^{\rm eff}$ (Eq.\ (\ref{eq:cagammaZ_1L})), $c_{aZZ}^{\rm eff}$ (Eq.\ (\ref{eq:caZZ_1L})) and $c_{aWW}^{\rm eff}$ (Eq.\ (\ref{eq:caWW_1L})), respectively, can be obtained from the general expressions for $A^{\text{f}}$, $B^{\text{f}}$ and $C^{\text{f}}$ for $Q_\text{f} = \frac{2}{3}$, $N_C = 3$ 
and $T_{3,f} = \frac{1}{2}$:
\begin{flalign}
A^{\text{f}}= 
 Q_\text{f} N_C \left\{ 2 Q_\text{f} s_w^2 + \frac{4 \left( T_{3,\text{f}} - 2 Q_\text{f} s_w^2 \right) m_\text{f}^2}{p^2 - M_Z^2} \left(f\left( \frac{4 m_\text{f}^2}{p^2} \right)^2 - f\left( \frac{4 m_\text{f}^2}{M_Z^2} \right)^2  \right) \right\} 
 \label{eq:Af}
\end{flalign}
with the function $f$ as defined in Eq.\ (\ref{eq:B1}).
\begin{equation}
\begin{aligned}
B^{\text{f}}=&
    - N_C \Bigg\{ Q_\mf^2 s_w^4 +  T^2_{3,f} \frac{2 m_\text{f}^2}{(4 M_Z^2 - p^2)} \left( \mathcal{DB} (p^2 , m_\mf , m_\mf ) - \mathcal{DB} (M_Z^2 , m_\mf , m_\mf )  \right) + \\
    & + \frac{2 m_\text{f}^2}{(4 M_Z^2 - p^2)} \left[M_Z^2 (T_{3,\mf} - 2 Q_\mf s_w^2)^2 + p^2 Q_\mf s_w^2 (T_{3,\mf} -  Q_\mf s_w^2) \right] \mathcal{C}\left(M_Z^2,M_Z^2,p^2,m_\text{f},m_\text{f},m_\text{f} \right) \Bigg\} \,
    \label{eq:Bf}
\end{aligned}\end{equation}
\begin{equation}
\begin{aligned}
C^{\text{f}} = 
 &   - N_C \Bigg\{ \frac{m_\text{f}^2}{4 (4M_W^2-p^2)}\left(  \frac{m_\text{f}^2-m_\text{f'}^2}{M_W^2} - 1 \right) \log \left( \frac{m_\text{f}^2}{m_\text{f'}^2} \right) \\
    &+ \frac{ m_\text{f}^2}{2(4M_W^2 - p^2)} \left( \mathcal{DB} (p^2 , m_\mf , m_\mf ) -  \mathcal{DB} (M_W^2 , m_\mf , m_\mfp ) \right) \\
    & + \frac{m_\text{f}^2 (M_W^2 - m_\text{f}^2 + m_\text{f'}^2)}{2(4M_W^2 - p^2)}  \mathcal{C}\left(M_W^2,M_W^2,p^2,m_\text{f},m_\text{f'},m_\text{f} \right) \Bigg\}\,.  
    \label{eq:Cf}
\end{aligned}\end{equation}
 The disc function $\mathcal{DB}$ is defined as
 \begin{equation} \label{functionDB}
\mathcal{DB} (p^2 , m_1 , m_2 ) \equiv \frac{ \sqrt{\rho(p^2,m_1^2,m_2^2)}}{p^2} \log \left( \frac{m_1^2 + m_2^2 - p^2 + \sqrt{\rho(p^2,m_1^2,m_2^2)}}{2 m_1 m_2} \right) \,,
\end{equation}
where $\rho$ denotes the K\"{a}ll\'en function
\begin{equation}\label{functionrho}
\rho(a,b,c) \equiv a^4 + b^4 +c^4 -2 a^2 b^2 - 2 b^2 c^2 - 2 c^2 a^2\,.
\end{equation} 
$\mathcal{C}(q_1^2 , q_2^2 , p^2 , m_1 , m_2 , m_3)$ is the $\text{C}_0$ Passarino-Veltman function~\cite{Passarino:1978jh}, defined as
\begin{equation}
\begin{aligned}\label{functionC}
& \mathcal{C} (q_1^2 , q_2^2 , p^2 , m_1 , m_2 , m_3) \equiv \\
& \int_0^1 \text{d}x \int_0^x \text{d}y \frac{1}{(x-y)y q_1^2 - (x-y) (x-1) q_2^2 -y (x-1) p^2 -y m_1^2 - (x-y) m_2^2 + (x-1) m_3^2}.
 \end{aligned} 
\end{equation}
Finally, the loop function  $A_{1/2}^a(\tau)$ defined in Sec.~\ref{sec:models} is given by
\begin{equation}
A_{1/2}^a(\tau) = \frac{2}{\tau} f(\tau), \quad 
f(\tau) = 
\begin{cases}
\arcsin^2\left( \sqrt{1/\tau} \right), & \tau \geq 1, \\
-\frac{1}{4} \left[ \log\left( \frac{1 + \sqrt{1 - \tau}}{1 - \sqrt{1 - \tau}} \right) - i\pi \right]^2, & \tau < 1.
\end{cases}
\label{eq:A12}
\end{equation}

\bibliographystyle{JHEP}
\bibliography{axionsLHC}

\providecommand{\href}[2]{#2}\begingroup\raggedright\begin{thebibliography}{10}

\bibitem{Peccei:1977hh}
R.~D. Peccei and H.~R. Quinn, {\it {CP Conservation in the Presence of Instantons}},  {\em Phys. Rev. Lett.} {\bf 38} (1977) 1440--1443.

\bibitem{Peccei:1977ur}
R.~D. Peccei and H.~R. Quinn, {\it {Constraints Imposed by CP Conservation in the Presence of Instantons}},  {\em Phys. Rev. D} {\bf 16} (1977) 1791--1797.

\bibitem{Wilczek:1977pj}
F.~Wilczek, {\it {Problem of Strong $P$ and $T$ Invariance in the Presence of Instantons}},  {\em Phys. Rev. Lett.} {\bf 40} (1978) 279--282.

\bibitem{Weinberg:1977ma}
S.~Weinberg, {\it {A New Light Boson?}},  {\em Phys. Rev. Lett.} {\bf 40} (1978) 223--226.

\bibitem{Choi:2020rgn}
K.~Choi, S.~H. Im, and C.~Sub~Shin, {\it {Recent Progress in the Physics of Axions and Axion-Like Particles}},  {\em Ann. Rev. Nucl. Part. Sci.} {\bf 71} (2021) 225--252, [\href{http://arxiv.org/abs/2012.05029}{{\tt arXiv:2012.05029}}].

\bibitem{AxionLimits}
C.~O’Hare, ``cajohare/axionlimits: Axionlimits.'' \url{https://cajohare.github.io/AxionLimits/}, July, 2020.

\bibitem{Mimasu:2014nea}
K.~Mimasu and V.~Sanz, {\it {ALPs at Colliders}},  {\em JHEP} {\bf 06} (2015) 173, [\href{http://arxiv.org/abs/1409.4792}{{\tt arXiv:1409.4792}}].

\bibitem{Jaeckel:2015jla}
J.~Jaeckel and M.~Spannowsky, {\it {Probing MeV to 90 GeV axion-like particles with LEP and LHC}},  {\em Phys. Lett. B} {\bf 753} (2016) 482--487, [\href{http://arxiv.org/abs/1509.00476}{{\tt arXiv:1509.00476}}].

\bibitem{Brivio:2017ije}
I.~Brivio, M.~B. Gavela, L.~Merlo, K.~Mimasu, J.~M. No, R.~del Rey, and V.~Sanz, {\it {ALPs Effective Field Theory and Collider Signatures}},  {\em Eur. Phys. J. C} {\bf 77} (2017), no.~8 572, [\href{http://arxiv.org/abs/1701.05379}{{\tt arXiv:1701.05379}}].

\bibitem{Bauer:2017ris}
M.~Bauer, M.~Neubert, and A.~Thamm, {\it {Collider Probes of Axion-Like Particles}},  {\em JHEP} {\bf 12} (2017) 044, [\href{http://arxiv.org/abs/1708.00443}{{\tt arXiv:1708.00443}}].

\bibitem{Craig:2018kne}
N.~Craig, A.~Hook, and S.~Kasko, {\it {The Photophobic ALP}},  {\em JHEP} {\bf 09} (2018) 028, [\href{http://arxiv.org/abs/1805.06538}{{\tt arXiv:1805.06538}}].

\bibitem{Esser:2023fdo}
F.~Esser, M.~Madigan, V.~Sanz, and M.~Ubiali, {\it {On the coupling of axion-like particles to the top quark}},  {\em JHEP} {\bf 09} (2023) 063, [\href{http://arxiv.org/abs/2303.17634}{{\tt arXiv:2303.17634}}].

\bibitem{Bisal:2025jwv}
S.~Bisal, {\it {Constraining ALP-Top Interaction from the Chromoelectric Dipole Moment of the Top Quark}},  \href{http://arxiv.org/abs/2507.12570}{{\tt arXiv:2507.12570}}.

\bibitem{Esser:2024pnc}
F.~Esser, M.~Madigan, A.~Salas-Bernardez, V.~Sanz, and M.~Ubiali, {\it {Di-Higgs production via axion-like particles}},  {\em JHEP} {\bf 10} (2024) 164, [\href{http://arxiv.org/abs/2404.08062}{{\tt arXiv:2404.08062}}].

\bibitem{Gavela:2019cmq}
M.~B. Gavela, J.~M. No, V.~Sanz, and J.~F. de~Troc\'oniz, {\it {Nonresonant Searches for Axionlike Particles at the LHC}},  {\em Phys. Rev. Lett.} {\bf 124} (2020), no.~5 051802, [\href{http://arxiv.org/abs/1905.12953}{{\tt arXiv:1905.12953}}].

\bibitem{No:2015bsn}
J.~M. No, V.~Sanz, and J.~Setford, {\it {See-saw composite Higgs model at the LHC: Linking naturalness to the 750 GeV diphoton resonance}},  {\em Phys. Rev. D} {\bf 93} (2016), no.~9 095010, [\href{http://arxiv.org/abs/1512.05700}{{\tt arXiv:1512.05700}}].

\bibitem{Carra:2021ycg}
S.~Carra, V.~Goumarre, R.~Gupta, S.~Heim, B.~Heinemann, J.~Kuechler, F.~Meloni, P.~Quilez, and Y.-C. Yap, {\it {Constraining off-shell production of axionlike particles with Z\ensuremath{\gamma} and WW differential cross-section measurements}},  {\em Phys. Rev. D} {\bf 104} (2021), no.~9 092005, [\href{http://arxiv.org/abs/2106.10085}{{\tt arXiv:2106.10085}}].

\bibitem{Alimena:2019zri}
J.~Alimena et~al., {\it {Searching for long-lived particles beyond the Standard Model at the Large Hadron Collider}},  {\em J. Phys. G} {\bf 47} (2020), no.~9 090501, [\href{http://arxiv.org/abs/1903.04497}{{\tt arXiv:1903.04497}}].

\bibitem{CMS:2018erd}
{\bf CMS} Collaboration, A.~M. Sirunyan et~al., {\it {Evidence for light-by-light scattering and searches for axion-like particles in ultraperipheral PbPb collisions at $\sqrt{s_\mathrm{NN}} =$ 5.02 TeV}},  {\em Phys. Lett. B} {\bf 797} (2019) 134826, [\href{http://arxiv.org/abs/1810.04602}{{\tt arXiv:1810.04602}}].

\bibitem{ATLAS:2022abz}
{\bf ATLAS} Collaboration, G.~Aad et~al., {\it {Search for boosted diphoton resonances in the 10 to 70 GeV mass range using 138 ifb of 13 TeV pp collisions with the ATLAS detector}},  {\em JHEP} {\bf 07} (2023) 155, [\href{http://arxiv.org/abs/2211.04172}{{\tt arXiv:2211.04172}}].

\bibitem{ATLAS:2023ofo}
{\bf ATLAS} Collaboration, G.~Aad et~al., {\it {Search for a new pseudoscalar decaying into a pair of muons in events with a top-quark pair at s=13{\,}{\,}TeV with the ATLAS detector}},  {\em Phys. Rev. D} {\bf 108} (2023), no.~9 092007, [\href{http://arxiv.org/abs/2304.14247}{{\tt arXiv:2304.14247}}].

\bibitem{ATLAS:2023ian}
{\bf ATLAS} Collaboration, G.~Aad et~al., {\it {Search for short- and long-lived axion-like particles in $H\rightarrow a a \rightarrow 4\gamma $ decays with the ATLAS experiment at the LHC}},  {\em Eur. Phys. J. C} {\bf 84} (2024), no.~7 742, [\href{http://arxiv.org/abs/2312.03306}{{\tt arXiv:2312.03306}}].

\bibitem{ATLAS:2023zfc}
{\bf ATLAS} Collaboration, G.~Aad et~al., {\it {Search for an axion-like particle with forward proton scattering in association with photon pairs at ATLAS}},  {\em JHEP} {\bf 07} (2023) 234, [\href{http://arxiv.org/abs/2304.10953}{{\tt arXiv:2304.10953}}].

\bibitem{ATLAS:2023etl}
{\bf ATLAS} Collaboration, G.~Aad et~al., {\it {Search for the decay of the Higgs boson to a $Z$ boson and a light pseudoscalar particle decaying to two photons}},  {\em Phys. Lett. B} {\bf 850} (2024) 138536, [\href{http://arxiv.org/abs/2312.01942}{{\tt arXiv:2312.01942}}].

\bibitem{ATLAS:2024vpj}
{\bf ATLAS} Collaboration, G.~Aad et~al., {\it {Search for decays of the Higgs boson into a pair of pseudoscalar particles decaying into bb{\textasciimacron}{\ensuremath{\tau}}+{\ensuremath{\tau}}- using pp collisions at s=13{\,}{\,}TeV with the ATLAS detector}},  {\em Phys. Rev. D} {\bf 110} (2024), no.~5 052013, [\href{http://arxiv.org/abs/2407.01335}{{\tt arXiv:2407.01335}}].

\bibitem{CMS:2024ulc}
{\bf CMS} Collaboration, A.~Tumasyan et~al., {\it {Search for a scalar or pseudoscalar dilepton resonance produced in association with a massive vector boson or top quark-antiquark pair in multilepton events at s=13{\,}{\,}TeV}},  {\em Phys. Rev. D} {\bf 110} (2024), no.~1 012013, [\href{http://arxiv.org/abs/2402.11098}{{\tt arXiv:2402.11098}}].

\bibitem{Arganda:2018cuz}
E.~Arganda, A.~D. Medina, N.~I. Mileo, R.~A. Morales, and A.~Szynkman, {\it {Constraining R-axion models through dijet searches at the LHC}},  {\em Phys. Lett. B} {\bf 789} (2019) 575--581, [\href{http://arxiv.org/abs/1808.01292}{{\tt arXiv:1808.01292}}].

\bibitem{Haghighat:2020nuh}
G.~Haghighat, D.~Haji~Raissi, and M.~Mohammadi~Najafabadi, {\it {New collider searches for axionlike particles coupling to gluons}},  {\em Phys. Rev. D} {\bf 102} (2020), no.~11 115010, [\href{http://arxiv.org/abs/2006.05302}{{\tt arXiv:2006.05302}}].

\bibitem{Redi:2011zi}
M.~Redi and A.~Weiler, {\it {Flavor and CP Invariant Composite Higgs Models}},  {\em JHEP} {\bf 11} (2011) 108, [\href{http://arxiv.org/abs/1106.6357}{{\tt arXiv:1106.6357}}].

\bibitem{Matsedonskyi:2012ym}
O.~Matsedonskyi, G.~Panico, and A.~Wulzer, {\it {Light Top Partners for a Light Composite Higgs}},  {\em JHEP} {\bf 01} (2013) 164, [\href{http://arxiv.org/abs/1204.6333}{{\tt arXiv:1204.6333}}].

\bibitem{Pomarol:2012qf}
A.~Pomarol and F.~Riva, {\it {The Composite Higgs and Light Resonance Connection}},  {\em JHEP} {\bf 08} (2012) 135, [\href{http://arxiv.org/abs/1205.6434}{{\tt arXiv:1205.6434}}].

\bibitem{Alloul:2013bka}
A.~Alloul, N.~D. Christensen, C.~Degrande, C.~Duhr, and B.~Fuks, {\it {FeynRules 2.0 - A complete toolbox for tree-level phenomenology}},  {\em Comput. Phys. Commun.} {\bf 185} (2014) 2250--2300, [\href{http://arxiv.org/abs/1310.1921}{{\tt arXiv:1310.1921}}].

\bibitem{Degrande:2011ua}
C.~Degrande, C.~Duhr, B.~Fuks, D.~Grellscheid, O.~Mattelaer, and T.~Reiter, {\it {UFO - The Universal FeynRules Output}},  {\em Comput. Phys. Commun.} {\bf 183} (2012) 1201--1214, [\href{http://arxiv.org/abs/1108.2040}{{\tt arXiv:1108.2040}}].

\bibitem{Alwall:2014hca}
J.~Alwall, R.~Frederix, S.~Frixione, V.~Hirschi, F.~Maltoni, O.~Mattelaer, H.~S. Shao, T.~Stelzer, P.~Torrielli, and M.~Zaro, {\it {The automated computation of tree-level and next-to-leading order differential cross sections, and their matching to parton shower simulations}},  {\em JHEP} {\bf 07} (2014) 079, [\href{http://arxiv.org/abs/1405.0301}{{\tt arXiv:1405.0301}}].

\bibitem{Buckley:2010ar}
A.~Buckley, J.~Butterworth, D.~Grellscheid, H.~Hoeth, L.~Lonnblad, J.~Monk, H.~Schulz, and F.~Siegert, {\it {Rivet user manual}},  {\em Comput. Phys. Commun.} {\bf 184} (2013) 2803--2819, [\href{http://arxiv.org/abs/1003.0694}{{\tt arXiv:1003.0694}}].

\bibitem{Bierlich:2024vqo}
C.~Bierlich, A.~Buckley, J.~M. Butterworth, C.~Gutschow, L.~Lonnblad, T.~Procter, P.~Richardson, and Y.~Yeh, {\it {Robust independent validation of experiment and theory: Rivet version 4 release note}},  {\em SciPost Phys. Codeb.} {\bf 36} (2024) 1, [\href{http://arxiv.org/abs/2404.15984}{{\tt arXiv:2404.15984}}].

\bibitem{Butterworth:2016sqg}
J.~M. Butterworth, D.~Grellscheid, M.~Kr\"amer, B.~Sarrazin, and D.~Yallup, {\it {Constraining new physics with collider measurements of Standard Model signatures}},  {\em JHEP} {\bf 03} (2017) 078, [\href{http://arxiv.org/abs/1606.05296}{{\tt arXiv:1606.05296}}].

\bibitem{Buckley:2021neu}
A.~Buckley et~al., {\it {Testing new physics models with global comparisons to collider measurements: the Contur toolkit}},  {\em SciPost Phys. Core} {\bf 4} (2021) 013, [\href{http://arxiv.org/abs/2102.04377}{{\tt arXiv:2102.04377}}].

\bibitem{CONTUR:2025yis}
{\bf CONTUR} Collaboration, A.~Buckley, J.~Butterworth, J.~Egan, C.~Gutschow, S.~Jeon, M.~Habedank, T.~Procter, P.~Wang, Y.~Yeh, and L.~Yue, {\it {Constraints On New Theories Using Rivet : CONTUR version 3 release note}},  \href{http://arxiv.org/abs/2505.09272}{{\tt arXiv:2505.09272}}.

\bibitem{Bonilla:2022qgm}
J.~Bonilla, A.~de~Giorgi, B.~Gavela, L.~Merlo, and M.~Ramos, {\it {The cost of an ALP solution to the neutral $B$-anomalies}},  \href{http://arxiv.org/abs/2209.11247}{{\tt arXiv:2209.11247}}.

\bibitem{Carmona:2022jid}
A.~Carmona, F.~Elahi, C.~Scherb, and P.~Schwaller, {\it {The ALPs from the top: searching for long lived axion-like particles from exotic top decays}},  {\em JHEP} {\bf 07} (2022) 122, [\href{http://arxiv.org/abs/2202.09371}{{\tt arXiv:2202.09371}}].

\bibitem{Carmona:2021seb}
A.~Carmona, C.~Scherb, and P.~Schwaller, {\it {Charming ALPs}},  {\em JHEP} {\bf 08} (2021) 121, [\href{http://arxiv.org/abs/2101.07803}{{\tt arXiv:2101.07803}}].

\bibitem{Chala:2020wvs}
M.~Chala, G.~Guedes, M.~Ramos, and J.~Santiago, {\it {Running in the ALPs}},  {\em Eur. Phys. J. C} {\bf 81} (2021), no.~2 181, [\href{http://arxiv.org/abs/2012.09017}{{\tt arXiv:2012.09017}}].

\bibitem{Bauer:2019gfk}
M.~Bauer, M.~Neubert, S.~Renner, M.~Schnubel, and A.~Thamm, {\it {Axionlike Particles, Lepton-Flavor Violation, and a New Explanation of $a_\mu$ and $a_e$}},  {\em Phys. Rev. Lett.} {\bf 124} (2020), no.~21 211803, [\href{http://arxiv.org/abs/1908.00008}{{\tt arXiv:1908.00008}}].

\bibitem{Bauer:2021mvw}
M.~Bauer, M.~Neubert, S.~Renner, M.~Schnubel, and A.~Thamm, {\it {Flavor probes of axion-like particles}},  {\em JHEP} {\bf 09} (2022) 056, [\href{http://arxiv.org/abs/2110.10698}{{\tt arXiv:2110.10698}}].

\bibitem{Bonilla:2021ufe}
J.~Bonilla, I.~Brivio, M.~B. Gavela, and V.~Sanz, {\it {One-loop corrections to ALP couplings}},  {\em JHEP} {\bf 11} (2021) 168, [\href{http://arxiv.org/abs/2107.11392}{{\tt arXiv:2107.11392}}].

\bibitem{Gripaios:2009pe}
B.~Gripaios, A.~Pomarol, F.~Riva, and J.~Serra, {\it {Beyond the Minimal Composite Higgs Model}},  {\em JHEP} {\bf 04} (2009) 070, [\href{http://arxiv.org/abs/0902.1483}{{\tt arXiv:0902.1483}}].

\bibitem{Contino:2011np}
R.~Contino, D.~Marzocca, D.~Pappadopulo, and R.~Rattazzi, {\it {On the effect of resonances in composite Higgs phenomenology}},  {\em JHEP} {\bf 10} (2011) 081, [\href{http://arxiv.org/abs/1109.1570}{{\tt arXiv:1109.1570}}].

\bibitem{Ferretti:2013kya}
G.~Ferretti and D.~Karateev, {\it {Fermionic UV completions of Composite Higgs models}},  {\em JHEP} {\bf 03} (2014) 077, [\href{http://arxiv.org/abs/1312.5330}{{\tt arXiv:1312.5330}}].

\bibitem{Sanz:2015sua}
V.~Sanz and J.~Setford, {\it {Composite Higgses with seesaw EWSB}},  {\em JHEP} {\bf 12} (2015) 154, [\href{http://arxiv.org/abs/1508.06133}{{\tt arXiv:1508.06133}}].

\bibitem{Belvedere:2024wzg}
A.~Belvedere, C.~Englert, R.~Kogler, and M.~Spannowsky, {\it {Dispelling the $\sqrt{\mathcal {L}} $ myth for the High-Luminosity LHC}},  {\em Eur. Phys. J. C} {\bf 84} (2024), no.~7 715, [\href{http://arxiv.org/abs/2402.07985}{{\tt arXiv:2402.07985}}].

\bibitem{CMS:2021vhb}
{\bf CMS} Collaboration, A.~Tumasyan et~al., {\it {Measurement of differential $t \bar t$ production cross sections in the full kinematic range using lepton+jets events from proton-proton collisions at $\sqrt {s}$ = 13\,\,TeV}},  {\em Phys. Rev. D} {\bf 104} (2021), no.~9 092013, [\href{http://arxiv.org/abs/2108.02803}{{\tt arXiv:2108.02803}}].

\bibitem{ATLAS:2017cez}
{\bf ATLAS} Collaboration, M.~Aaboud et~al., {\it {Measurements of top-quark pair differential cross-sections in the lepton+jets channel in $pp$ collisions at $\sqrt{s}=13$ TeV using the ATLAS detector}},  {\em JHEP} {\bf 11} (2017) 191, [\href{http://arxiv.org/abs/1708.00727}{{\tt arXiv:1708.00727}}].

\bibitem{ATLAS:2019hxz}
{\bf ATLAS} Collaboration, G.~Aad et~al., {\it {Measurements of top-quark pair differential and double-differential cross-sections in the $\ell$+jets channel with $pp$ collisions at $\sqrt{s}=13$ TeV using the ATLAS detector}},  {\em Eur. Phys. J. C} {\bf 79} (2019), no.~12 1028, [\href{http://arxiv.org/abs/1908.07305}{{\tt arXiv:1908.07305}}]. [Erratum: Eur.Phys.J.C 80, 1092 (2020)].

\bibitem{ATLAS:2015mip}
{\bf ATLAS} Collaboration, G.~Aad et~al., {\it {Measurement of the differential cross-section of highly boosted top quarks as a function of their transverse momentum in $\sqrt{s}$ = 8 TeV proton-proton collisions using the ATLAS detector}},  {\em Phys. Rev. D} {\bf 93} (2016), no.~3 032009, [\href{http://arxiv.org/abs/1510.03818}{{\tt arXiv:1510.03818}}].

\bibitem{ATLAS:2017luz}
{\bf ATLAS} Collaboration, M.~Aaboud et~al., {\it {Measurements of electroweak $Wjj$ production and constraints on anomalous gauge couplings with the ATLAS detector}},  {\em Eur. Phys. J. C} {\bf 77} (2017), no.~7 474, [\href{http://arxiv.org/abs/1703.04362}{{\tt arXiv:1703.04362}}].

\bibitem{ATLAS:2022xfj}
{\bf ATLAS} Collaboration, G.~Aad et~al., {\it {Measurements of differential cross-sections in top-quark pair events with a high transverse momentum top quark and limits on beyond the Standard Model contributions to top-quark pair production with the ATLAS detector at $ \sqrt{s} $ = 13 TeV}},  {\em JHEP} {\bf 06} (2022) 063, [\href{http://arxiv.org/abs/2202.12134}{{\tt arXiv:2202.12134}}].

\bibitem{ATLAS:2017irc}
{\bf ATLAS} Collaboration, M.~Aaboud et~al., {\it {Measurement of differential cross sections and $W^+/W^-$ cross-section ratios for $W$ boson production in association with jets at $\sqrt{s}=8$ TeV with the ATLAS detector}},  {\em JHEP} {\bf 05} (2018) 077, [\href{http://arxiv.org/abs/1711.03296}{{\tt arXiv:1711.03296}}]. [Erratum: JHEP 10, 048 (2020)].

\bibitem{CMS:2016oae}
{\bf CMS} Collaboration, V.~Khachatryan et~al., {\it {Measurement of differential cross sections for top quark pair production using the lepton+jets final state in proton-proton collisions at 13 TeV}},  {\em Phys. Rev. D} {\bf 95} (2017), no.~9 092001, [\href{http://arxiv.org/abs/1610.04191}{{\tt arXiv:1610.04191}}].

\bibitem{CMS:2018tdx}
{\bf CMS} Collaboration, A.~M. Sirunyan et~al., {\it {Measurements of differential cross sections of top quark pair production as a function of kinematic event variables in proton-proton collisions at $ \sqrt{s}=13 $ TeV}},  {\em JHEP} {\bf 06} (2018) 002, [\href{http://arxiv.org/abs/1803.03991}{{\tt arXiv:1803.03991}}].

\bibitem{CMS:2018htd}
{\bf CMS} Collaboration, A.~M. Sirunyan et~al., {\it {Measurement of differential cross sections for the production of top quark pairs and of additional jets in lepton+jets events from pp collisions at $\sqrt{s} =$ 13 TeV}},  {\em Phys. Rev. D} {\bf 97} (2018), no.~11 112003, [\href{http://arxiv.org/abs/1803.08856}{{\tt arXiv:1803.08856}}].

\bibitem{ATLAS:2015lsn}
{\bf ATLAS} Collaboration, G.~Aad et~al., {\it {Measurements of top-quark pair differential cross-sections in the lepton+jets channel in $pp$ collisions at $\sqrt{s}=8$ TeV using the ATLAS detector}},  {\em Eur. Phys. J. C} {\bf 76} (2016), no.~10 538, [\href{http://arxiv.org/abs/1511.04716}{{\tt arXiv:1511.04716}}].

\bibitem{ATLAS:2017xqp}
{\bf ATLAS} Collaboration, M.~Aaboud et~al., {\it {Measurement of the cross section for isolated-photon plus jet production in $pp$ collisions at $\sqrt s=13$ TeV using the ATLAS detector}},  {\em Phys. Lett. B} {\bf 780} (2018) 578--602, [\href{http://arxiv.org/abs/1801.00112}{{\tt arXiv:1801.00112}}].

\bibitem{ATLAS:2024vqf}
{\bf ATLAS} Collaboration, G.~Aad et~al., {\it {Differential cross-sections for events with missing transverse momentum and jets measured with the ATLAS detector in 13 TeV proton-proton collisions}},  {\em JHEP} {\bf 08} (2024) 223, [\href{http://arxiv.org/abs/2403.02793}{{\tt arXiv:2403.02793}}].

\bibitem{ATLAS:2019iaa}
{\bf ATLAS} Collaboration, G.~Aad et~al., {\it {Measurement of isolated-photon plus two-jet production in $pp$ collisions at $\sqrt s=13$ TeV with the ATLAS detector}},  {\em JHEP} {\bf 03} (2020) 179, [\href{http://arxiv.org/abs/1912.09866}{{\tt arXiv:1912.09866}}].

\bibitem{ATLAS:2018orx}
{\bf ATLAS} Collaboration, M.~Aaboud et~al., {\it {Measurements of $t\bar{t}$ differential cross-sections of highly boosted top quarks decaying to all-hadronic final states in $pp$ collisions at $\sqrt{s}=13\,$ TeV using the ATLAS detector}},  {\em Phys. Rev. D} {\bf 98} (2018), no.~1 012003, [\href{http://arxiv.org/abs/1801.02052}{{\tt arXiv:1801.02052}}].

\bibitem{ATLAS:2020ccu}
{\bf ATLAS} Collaboration, G.~Aad et~al., {\it {Measurements of top-quark pair single- and double-differential cross-sections in the all-hadronic channel in $pp$ collisions at $\sqrt{s}=13~\textrm{TeV}$ using the ATLAS detector}},  {\em JHEP} {\bf 01} (2021) 033, [\href{http://arxiv.org/abs/2006.09274}{{\tt arXiv:2006.09274}}].

\bibitem{CMS:2019fak}
{\bf CMS} Collaboration, A.~M. Sirunyan et~al., {\it {Measurement of the Jet Mass Distribution and Top Quark Mass in Hadronic Decays of Boosted Top Quarks in $pp$ Collisions at $\sqrt{s} =$ TeV}},  {\em Phys. Rev. Lett.} {\bf 124} (2020), no.~20 202001, [\href{http://arxiv.org/abs/1911.03800}{{\tt arXiv:1911.03800}}].

\bibitem{ATLAS:2022mlu}
{\bf ATLAS} Collaboration, G.~Aad et~al., {\it {Differential $ t\overline{t} $ cross-section measurements using boosted top quarks in the all-hadronic final state with 139 fb$^{-1}$ of ATLAS data}},  {\em JHEP} {\bf 04} (2023) 080, [\href{http://arxiv.org/abs/2205.02817}{{\tt arXiv:2205.02817}}].

\bibitem{CMS:2019eih}
{\bf CMS} Collaboration, A.~M. Sirunyan et~al., {\it {Measurement of the $\mathrm{t\bar{t}}\mathrm{b\bar{b}}$ production cross section in the all-jet final state in pp collisions at $\sqrt{s} =$ 13 TeV}},  {\em Phys. Lett. B} {\bf 803} (2020) 135285, [\href{http://arxiv.org/abs/1909.05306}{{\tt arXiv:1909.05306}}].

\bibitem{ATLAS:2017txd}
{\bf ATLAS} Collaboration, M.~Aaboud et~al., {\it {Measurement of detector-corrected observables sensitive to the anomalous production of events with jets and large missing transverse momentum in $pp$ collisions at $\mathbf {\sqrt{s}=13}$ TeV using the ATLAS detector}},  {\em Eur. Phys. J. C} {\bf 77} (2017), no.~11 765, [\href{http://arxiv.org/abs/1707.03263}{{\tt arXiv:1707.03263}}].

\bibitem{CMS:2022woe}
{\bf CMS} Collaboration, A.~Tumasyan et~al., {\it {Observation of electroweak W+W{\ensuremath{-}} pair production in association with two jets in proton-proton collisions at s=13TeV}},  {\em Phys. Lett. B} {\bf 841} (2023) 137495, [\href{http://arxiv.org/abs/2205.05711}{{\tt arXiv:2205.05711}}].

\bibitem{CMS:2020mxy}
{\bf CMS} Collaboration, A.~M. Sirunyan et~al., {\it {W$^+$W$^-$ boson pair production in proton-proton collisions at $\sqrt{s} =$ 13 TeV}},  {\em Phys. Rev. D} {\bf 102} (2020), no.~9 092001, [\href{http://arxiv.org/abs/2009.00119}{{\tt arXiv:2009.00119}}].

\bibitem{ATLAS:2023gsl}
{\bf ATLAS} Collaboration, G.~Aad et~al., {\it {Inclusive and differential cross-sections for dilepton $ t\overline{t} $ production measured in $ \sqrt{s} $ = 13 TeV pp collisions with the ATLAS detector}},  {\em JHEP} {\bf 07} (2023) 141, [\href{http://arxiv.org/abs/2303.15340}{{\tt arXiv:2303.15340}}].

\bibitem{ATLAS:2019hau}
{\bf ATLAS} Collaboration, G.~Aad et~al., {\it {Measurement of the $t\bar{t}$ production cross-section and lepton differential distributions in $e\mu $ dilepton events from $pp$ collisions at $\sqrt{s}=13\,\text {TeV}$ with the ATLAS detector}},  {\em Eur. Phys. J. C} {\bf 80} (2020), no.~6 528, [\href{http://arxiv.org/abs/1910.08819}{{\tt arXiv:1910.08819}}].

\bibitem{ATLAS:2019ebv}
{\bf ATLAS} Collaboration, M.~Aaboud et~al., {\it {Searches for scalar leptoquarks and differential cross-section measurements in dilepton-dijet events in proton-proton collisions at a centre-of-mass energy of $\sqrt{s}$ = 13 TeV with the ATLAS experiment}},  {\em Eur. Phys. J. C} {\bf 79} (2019), no.~9 733, [\href{http://arxiv.org/abs/1902.00377}{{\tt arXiv:1902.00377}}].

\bibitem{ATLAS:2015xtc}
{\bf ATLAS} Collaboration, G.~Aad et~al., {\it {Measurement of four-jet differential cross sections in $\sqrt{s}=8$ TeV proton-proton collisions using the ATLAS detector}},  {\em JHEP} {\bf 12} (2015) 105, [\href{http://arxiv.org/abs/1509.07335}{{\tt arXiv:1509.07335}}].

\bibitem{ATLAS:2017ble}
{\bf ATLAS} Collaboration, M.~Aaboud et~al., {\it {Measurement of inclusive jet and dijet cross-sections in proton-proton collisions at $\sqrt{s}=13$ TeV with the ATLAS detector}},  {\em JHEP} {\bf 05} (2018) 195, [\href{http://arxiv.org/abs/1711.02692}{{\tt arXiv:1711.02692}}].

\bibitem{ATLAS:2017nei}
{\bf ATLAS} Collaboration, M.~Aaboud et~al., {\it {Measurement of the cross-section for electroweak production of dijets in association with a Z boson in pp collisions at $\sqrt {s}$ = 13 TeV with the ATLAS detector}},  {\em Phys. Lett. B} {\bf 775} (2017) 206--228, [\href{http://arxiv.org/abs/1709.10264}{{\tt arXiv:1709.10264}}].

\bibitem{ATLAS:2024tnr}
{\bf ATLAS} Collaboration, G.~Aad et~al., {\it {Measurements of the production cross-section for a Z boson in association with b- or c-jets in proton\textendash{}proton collisions at $\sqrt{s} = 13$ TeV with the ATLAS detector}},  {\em Eur. Phys. J. C} {\bf 84} (2024), no.~9 984, [\href{http://arxiv.org/abs/2403.15093}{{\tt arXiv:2403.15093}}].

\bibitem{ATLAS:2020juj}
{\bf ATLAS} Collaboration, G.~Aad et~al., {\it {Measurements of the production cross-section for a $Z$ boson in association with $b$-jets in proton-proton collisions at $\sqrt{s} = 13$ TeV with the ATLAS detector}},  {\em JHEP} {\bf 07} (2020) 044, [\href{http://arxiv.org/abs/2003.11960}{{\tt arXiv:2003.11960}}].

\bibitem{Altakach:2020ugg}
M.~M. Altakach, T.~Je{\v{z}}o, M.~Klasen, J.-N. Lang, and I.~Schienbein, {\it {Electroweak tt{\textasciimacron} hadroproduction in the presence of heavy Z' and W' bosons at NLO QCD in POWHEG}},  {\em Phys. Rev. D} {\bf 103} (2021), no.~11 115026, [\href{http://arxiv.org/abs/2012.14855}{{\tt arXiv:2012.14855}}].

\bibitem{Sjostrand:2014zea}
T.~Sj{\"o}strand, S.~Ask, J.~R. Christiansen, R.~Corke, N.~Desai, P.~Ilten, S.~Mrenna, S.~Prestel, C.~O. Rasmussen, and P.~Z. Skands, {\it {An introduction to PYTHIA 8.2}},  {\em Comput. Phys. Commun.} {\bf 191} (2015) 159--177, [\href{http://arxiv.org/abs/1410.3012}{{\tt arXiv:1410.3012}}].

\bibitem{Alioli:2010xd}
S.~Alioli, P.~Nason, C.~Oleari, and E.~Re, {\it {A general framework for implementing NLO calculations in shower Monte Carlo programs: the POWHEG BOX}},  {\em JHEP} {\bf 06} (2010) 043, [\href{http://arxiv.org/abs/1002.2581}{{\tt arXiv:1002.2581}}].

\bibitem{Bellm:2015jjp}
J.~Bellm et~al., {\it {Herwig 7.0/Herwig++ 3.0 release note}},  {\em Eur. Phys. J. C} {\bf 76} (2016), no.~4 196, [\href{http://arxiv.org/abs/1512.01178}{{\tt arXiv:1512.01178}}].

\bibitem{Frixione:2007vw}
S.~Frixione, P.~Nason, and C.~Oleari, {\it {Matching NLO QCD computations with Parton Shower simulations: the POWHEG method}},  {\em JHEP} {\bf 11} (2007) 070, [\href{http://arxiv.org/abs/0709.2092}{{\tt arXiv:0709.2092}}].

\bibitem{Chen:2019zmr}
X.~Chen, T.~Gehrmann, N.~Glover, M.~H{\"o}fer, and A.~Huss, {\it {Isolated photon and photon+jet production at NNLO QCD accuracy}},  {\em JHEP} {\bf 04} (2020) 166, [\href{http://arxiv.org/abs/1904.01044}{{\tt arXiv:1904.01044}}].

\bibitem{CMS:2018mdf}
{\bf CMS} Collaboration, A.~M. Sirunyan et~al., {\it {Measurement of differential cross sections for Z boson production in association with jets in proton-proton collisions at $\sqrt{s} =$ 13 TeV}},  {\em Eur. Phys. J. C} {\bf 78} (2018), no.~11 965, [\href{http://arxiv.org/abs/1804.05252}{{\tt arXiv:1804.05252}}].

\bibitem{CMS:2019raw}
{\bf CMS} Collaboration, A.~M. Sirunyan et~al., {\it {Measurements of differential Z boson production cross sections in proton-proton collisions at $ \sqrt{s} $ = 13 TeV}},  {\em JHEP} {\bf 12} (2019) 061, [\href{http://arxiv.org/abs/1909.04133}{{\tt arXiv:1909.04133}}].

\bibitem{CMS:2021wfx}
{\bf CMS} Collaboration, A.~Tumasyan et~al., {\it {Study of Z boson plus jets events using variables sensitive to double-parton scattering in pp collisions at 13 TeV}},  {\em JHEP} {\bf 10} (2021) 176, [\href{http://arxiv.org/abs/2105.14511}{{\tt arXiv:2105.14511}}].

\bibitem{CMS:2022ubq}
{\bf CMS} Collaboration, A.~Tumasyan et~al., {\it {Measurement of the mass dependence of the transverse momentum of lepton pairs in Drell-Yan production in proton-proton collisions at $\sqrt{s}$ = 13 TeV}},  {\em Eur. Phys. J. C} {\bf 83} (2023), no.~7 628, [\href{http://arxiv.org/abs/2205.04897}{{\tt arXiv:2205.04897}}].

\bibitem{Monni:2019whf}
P.~F. Monni, P.~Nason, E.~Re, M.~Wiesemann, and G.~Zanderighi, {\it {MiNNLO$_{PS}$: a new method to match NNLO QCD to parton showers}},  {\em JHEP} {\bf 05} (2020) 143, [\href{http://arxiv.org/abs/1908.06987}{{\tt arXiv:1908.06987}}]. [Erratum: JHEP 02, 031 (2022)].

\bibitem{Monni:2020nks}
P.~F. Monni, E.~Re, and M.~Wiesemann, {\it {MiNNLO$_{\text {PS}}$: optimizing $2\rightarrow 1$ hadronic processes}},  {\em Eur. Phys. J. C} {\bf 80} (2020), no.~11 1075, [\href{http://arxiv.org/abs/2006.04133}{{\tt arXiv:2006.04133}}].

\bibitem{Passarino:1978jh}
G.~Passarino and M.~Veltman, {\it One-loop corrections for $e^+e^-$ annihilation into $\mu^+\mu^-$ in the weinberg model},  {\em Nuclear Physics B} {\bf 160} (1979), no.~1 151--207.

\end{thebibliography}\endgroup

\end{document}